\documentclass[10pt,journal,doublecolumn]{IEEEtran}
\IEEEoverridecommandlockouts

\usepackage[pdftex]{graphicx}
\graphicspath{{Figures/}{../png/}{../jpg/}{../pdf/}}

\usepackage{epstopdf}

\usepackage{amsthm}
\usepackage{amssymb}
\usepackage{mathptmx}
\usepackage{color,soul}
\usepackage[cmex10]{amsmath}
\usepackage{amsfonts}
\usepackage{caption}
\usepackage{comment}
\usepackage{xcolor}
\usepackage[boxed,ruled,vlined,linesnumbered,commentsnumbered, noresetcount]{algorithm2e}
\usepackage{algorithmic}
\usepackage{setspace}
\usepackage{mathtools}
\usepackage{cite}
\usepackage{enumitem}

\newcommand{\figwidth}{0.9\columnwidth}

\makeatletter
\newcommand{\biggg}{\bBigg@{3}}
\newcommand{\vast}{\bBigg@{4}}
\newcommand{\Vast}{\bBigg@{5}}
\makeatother

\usepackage{threeparttable,placeins,makecell,stfloats,cite}
\usepackage{threeparttable,arydshln}
\usepackage{amsmath,amssymb}
\usepackage{amsfonts}
\usepackage{latexsym}
\usepackage{multirow}
\usepackage{caption}
\usepackage{rotating}
\usepackage{float}
\usepackage{url}
\usepackage{cite}
\usepackage{rotating}
\usepackage{float}
\usepackage[format=hang]{subfig}
\usepackage{url}
\usepackage{cite}
\usepackage{multirow}

\usepackage{booktabs}

\usepackage{amssymb}
\usepackage{stackengine}
\usepackage{scalerel}
\usepackage{xcolor}
\usepackage{graphicx}
 \newcommand\obigstar[1][0.7]{%
  \scalerel*{%
    \stackinset{c}{-.125pt}{c}{}{\scalebox{#1}{\color{white}{$\bigstar$}}}{%
      $\bigstar$}%
  }{\bigstar}
}
\usepackage{array}
\newcolumntype{P}[1]{>{\centering\arraybackslash}p{#1}}
\newcolumntype{C}[1]{>{\centering\arraybackslash}m{#1}}
\newcolumntype{R}[1]{>{\raggedleft\arraybackslash}m{#1}}
\newcolumntype{L}[1]{>{\raggedright\arraybackslash}m{#1}}

\usepackage[none]{hyphenat}
\usepackage{flushend}

\begin{document}


\title{6G for Vehicle-to-Everything (V2X) Communications: Enabling Technologies, Challenges, and Opportunities}


\author{Md. Noor-A-Rahim,~Zilong~Liu, Haeyoung Lee, M. Omar Khyam, Jianhua He, Dirk Pesch,  Klaus Moessner, Walid~Saad,  H.~Vincent~Poor
\thanks{M. Noor-A-Rahim and D. Pesch  are with the  School of Computer Science \& IT, University College Cork,  Ireland  (E-mail: {\tt \{m.rahim,d.pesch\}@cs.ucc.ie}).

Z. Liu and J. He are with the School of Computer Science and Electronics Engineering, University of Essex, UK (E-mail: {\tt \{zilong.liu, j.he\}@essex.ac.uk}).

H. Lee  is with the 5G \& 6G Innovation Centre (5GIC/6GIC), Institute for Communcation Systems (ICS), University of Surrey, U.K (E-mail: {\tt Haeyoung.Lee@surrey.ac.uk}).

M. O. Khyam is  with the  Defence Science and Technology Group (DSTG),  Australia.

K. Moessner is with the Institute for Communcation Systems (ICS), University of Surrey, U.K and the Faculty of Electronics and Information Technology,  University of Technology Chemnitz, Germany (E-mail: {\tt klaus.moessner@etit.tu-chemnitz.de}).

W. Saad is with Wireless@VT, the Bradley Department of Electrical and Computer Engineering, Virginia Tech, USA (E-mail: {\tt walids@vt.edu}).

H. V. Poor is with the  Department of Electrical and Computer Engineering, Princeton University, USA (E-mail: {\tt poor@princeton.edu}).
}
}
\maketitle

\thispagestyle{empty}
\pagestyle{empty}

\begin{abstract}
We are on the cusp of a new era of connected autonomous vehicles with unprecedented user experiences, tremendously improved road safety and air quality, highly diverse transportation environments and use cases, as well as a plethora of advanced applications. Realizing this grand vision requires a significantly enhanced vehicle-to-everything (V2X) communication network which should be extremely intelligent and capable of concurrently supporting hyper-fast, ultra-reliable, and low-latency massive information exchange. It is anticipated that the sixth-generation (6G) communication systems will fulfill these requirements of the next-generation V2X. In this article, we outline a series of key enabling technologies from a range of domains, such as new materials, algorithms, and system architectures. Aiming for truly intelligent transportation systems, we envision that machine learning will play an instrumental role for advanced vehicular communication and networking. To this end, we provide an overview on the recent advances of machine learning in 6G vehicular networks. To stimulate future research in this area, we discuss the strength, open challenges, maturity, and enhancing areas of these technologies.

\end{abstract}




\section{Introduction} \label{introduction}


\IEEEPARstart{I}{}n recent years, vehicle-to-everything (V2X) communication has attracted significant research interest by both academia and industry. As a key enabler for intelligent transportation systems (ITS), V2X encompasses a broad range of wireless technologies including vehicle-to-vehicle (V2V) communications, vehicle-to-infrastructure (V2I) communications, and vehicle-to-pedestrian (V2P) communications), as well as communications with  vulnerable road users (VRUs), and with cloud networks (V2N) \cite{Chen17}.  The grand vision is that V2X communications, supported by the sixth generation (6G) of wireless systems \cite{Saad2020}, will
 be an instrumental element of future connected autonomous vehicles. Furthermore, V2X communications will bring far-reaching and transformative benefits such as unprecedented user experience, tremendously improved road safety and air quality, diverse transportation applications and use cases, as well as a plethora of advanced applications. 


{So far, there have been two main technologies for V2X communications: 1) DSRC (Dedicated Short Range Communication)-based vehicular network and 2) the cellular-based vehicular network \cite{Naik19}. Standards laying the foundation of DSRC include IEEE 802.11p for Wireless Access in Vehicular Environments (WAVE) and IEEE 1609.1.4 for resource management, security, network service, and multichannel operation \cite{Kenney_2011}. For many years, DSRC was the only technology for V2X communication. In dense and high-mobility environments, however, DSRC suffers from major drawbacks such as limited coverage, low data rate, limited quality-of-service (QoS) guarantees, and unbounded channel access delay. Leveraging standard cellular technologies, 3GPP has been developing a cellular vehicular communications standard, known as C-V2X. C-V2X enables every vehicle to communicate with different entities of a V2X network (such as V2V, V2I, V2P, and V2N) \cite{Seo16}. In March 2017, 3GPP Release 14 proposed LTE V2X communication with two air interfaces: a wide area network LTE interface (LTE-Uu) and a direct communications interface (PC5, also known as LTE side-link). The LTE-Uu is responsible for vehicle to network (V2N) communication, while the LTE side-link is responsible for V2V and V2I communications, which may operate without support from cellular network infrastructure \cite{Sadio20}.} The main focus of Release~14 is to deliver data transport services for fundamental road safety services such as cooperative awareness messages (CAM), basic safety messages (BSM), or decentralized environmental notification messages (DENM).


\begin{figure*}[htbp]
	\centering
		\includegraphics[width=10cm, height = 4cm]{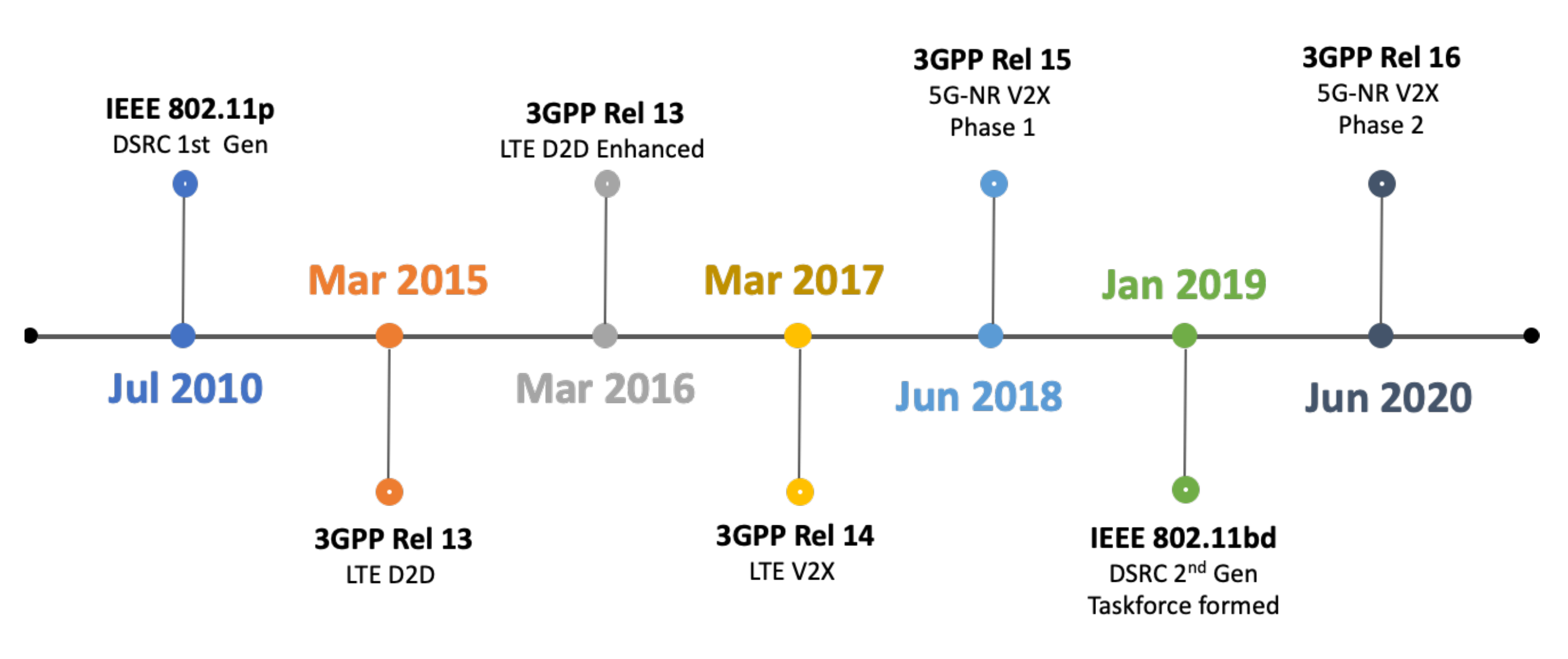}
	\caption{Evolution of V2X communications.}
	\label{fig:V2X_Evol}
\end{figure*}

In Release~15 (announced in 2018), 5G New Radio (5G NR) V2X technology was introduced  to support advanced V2X services such as vehicle platooning, advanced driver assistance, remote driving, and extended sensors  \cite{Release15v2x, bagheri20205g}. In addition, the performance of the PC5 interface has been enhanced in Release~15 (known as LTE-eV2X) in terms of higher reliability (employing transmit diversity), lower latency (with the aid of resource selection window reduction), and higher data rates (using carrier aggregation and higher order modulation e.g., 64-QAM), while retaining backward compatibility with Release~14 LTE-V2X. In 2020, 3GPP announced the second phase of 5G NR in Release~16, which aims at bringing enhanced ultra reliable low latency communication (URLLC) and higher throughput. Note that 3GPP is currently working on Release~17, which aims to provide architectural enhancements to support advanced V2X services. The evolution of V2X communications is summarized in Fig.~\ref{fig:V2X_Evol}.


{From an industry standpoint, there has been a major debate on which V2X technology should be adopted. Government regulations and public acceptance are two additional major factors which affect the real roll out of V2X technologies. Take the United States (US) as an example, during the Obama administration, a mandate was proposed by the National Highway Traffic Safety Administration to support DSRC in all new vehicles, but no progress has been made during the Trump administration. Although Europe and Japan are in favour of DSRC, C-V2X has received tremendous support from the US and China, mainly driven by large telecommunication companies such as Qualcomm and Huawei, respectively. As a result, automakers may need to investigate and accommodate different V2X technologies according to their market-shares in different jurisdictions as well as the inter-working of both DSRC and C-V2X \cite{Interworking2016}. For instance, although Toyota was the first global automaker to sell connected vehicles equipped with DSRC technology, it is also testing C-V2X in China, and both DSRC and C-V2X in Australia \cite{Toyota}. In 2019, Volkswagen, a German motor vehicle manufacturer, made a move to fully embrace DSRC \cite{Volkswagen}, while Ford in the US plans to enable all of its new vehicles to be able to ``Talk and Listen” by 2022 through 5G NR based C-V2X \cite{Ford}. Road safety and traffic efficiency are two important aspects for public acceptance. Albeit autonomous vehicles have been actively studied and tested by companies like Tesla, Uber, Waymo, and Toyota, the safety of driver-less cars remains of utmost concern for the public, especially after the pedestrian death caused by a self-driving Uber car in 2018 and the Tesla car crash in 2021. Therefore, it is pressing for V2X to evolve to facilitate advanced orchestrations of communication, sensing, learning and decision making. }


\begin{figure*}[htbp]
	\centering
        \includegraphics[width=5in]{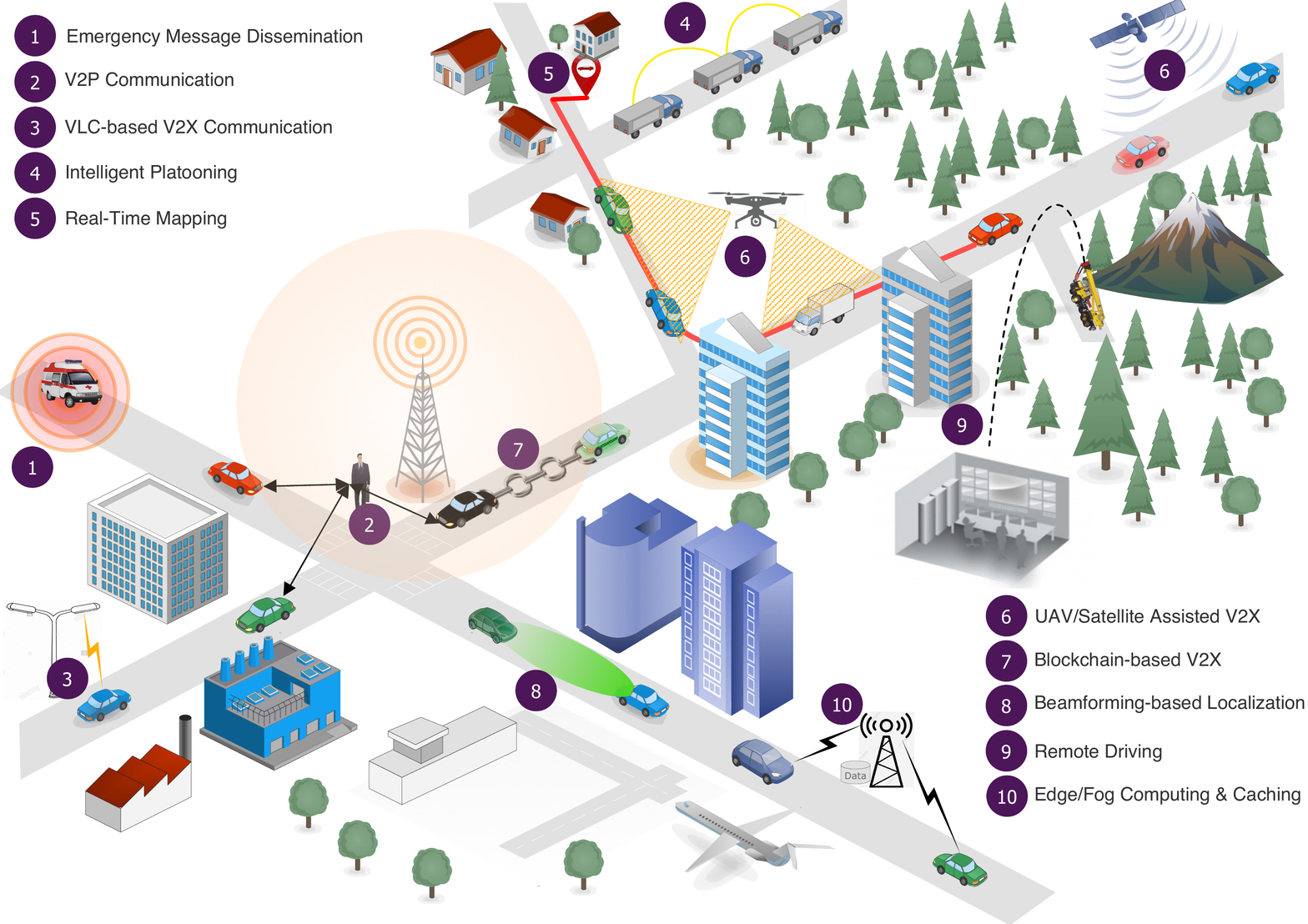}
	\caption{Overview of V2X communications.}
	\label{fig:V2X_Over}
\end{figure*}

\textbf{Contribution and Organization} The main contribution of this paper is a comprehensive overview of the scientific and technological advances that have the capability to shape future 6G vehicle-to-everything (6G-V2X) communications. {In contrast to the survey in \cite{V2X-Zhou-2020}, which focuses on the evolution of the two major V2X technologies (i.e., DSRC/802.11p based V2X and C-V2X) as well as big-data driven internet-of-vehicle (IoV) and cloud-based IoV, we take a forward-looking and inclusive approach from various perspectives (such as new materials, algorithms, and system structures), aiming to stimulate many forthcoming research activities for 6G-V2X and beyond. We also emphasize the instrumental role machine learning will play for advanced use cases in future vehicular communication networking.}

This article is organized as follows. We start our discourse in Section~\ref{sec:overview} by a high-level overview of 6G-V2X communications.  Revolutionary 6G-V2X technologies such as brain-vehicle interfacing, tactile communication, and terahertz communications, will be introduced in Section~\ref{sec:revtech}. In Section~\ref{sec:evotech}, we present major 6G-V2X technologies (e.g., integrated localization and communications, satellite/UAV aided V2X, integrated computing, etc) that have evolved over recent years and are going through further enhancements. The recent advances in machine learning for  6G  vehicular networks are summarized in Section~\ref{sec:Key_ML}. Finally, conclusions are drawn in Section~\ref{sec:conclu}.

\section{Overview of 6G-V2X Communications} \label{sec:overview}
This section first discusses why 6G-V2X is necessary  and then summarizes key technologies that will enable 6G-V2X.

\subsection{Why 6G-V2X?}
Although 5G-NR V2X offers improved performance with advanced services, its improved performance is achieved through investing more in spectral and hardware resources while inheriting the underlying mechanisms and system architectures of LTE-based V2X \cite{Tariq2020}. Meanwhile, it is anticipated that the number of autonomous vehicles will grow rapidly in the  future due to urbanization, increased living standards, and technological advancements. This will drive an explosive growth of communications devices and digital applications to enable intelligent autonomous vehicles. In addition, the rising demand for many emerging services in autonomous vehicles ranging from 3D displays that offer more depth and more natural viewing experience and free-floating, to holographic control display systems, to immersive entertainment, to improved in-car infotainment, will bring forth new communication challenges to the V2X network \cite{Nayak20,Akyildiz2020,Tariq2020,Xiao20,Xiaohu2020}. All these advances will drastically push the capacity limits of existing wireless networks, posing new scientific and technical challenges for vehicular networks in terms of data rate, latency, coverage, spectral/energy/cost efficiency, intelligence level, networking, and security, among others.

With  this vision in mind, 5G NR-based V2X networks may be unable to meet such a wide range of requirements and use cases. Moreover, while the concepts associated with ITS have been studied for many years, legacy V2X communication systems
can only provide limited intelligence.
Therefore, a significant paradigm shift away from traditional communication networks to more versatile and diversified network approaches is needed. It is anticipated that such a transformation will start from the recently proposed 6G wireless communication network, which aims to combine terrestrial and several non-terrestrial communication networks such as satellite and unmanned-aerial-vehicle (UAV) communication networks. This will enable genuinely intelligent and ubiquitous V2X systems with significantly enhanced reliability and security, extremely high data rates (e.g., Tbps), massive and hyper-fast wireless access (i.e., down to sub-milliseconds with billions of communications devices connected), as
well as much smarter, longer, and greener (energy-efficient) three-dimensional (3D) communication coverage \cite{Saad2020}. Because of the extremely heterogeneous network composition, diverse communication scenarios, and stringent service requirements,  new  techniques are needed to enable adaptive learning and intelligent decision making in future V2X networks. It is foreseen that 6G will work in conjunction with machine learning (ML) not only to unfold the full capability of the radio signals by evolving to intelligent and autonomous radios, but also to bring a series of new features such as enhanced context-awareness, self-aggregation, adaptive coordination, and self-configuration~\cite{Tariq2020}.

\begin{figure*}[htbp]
	\centering
        \includegraphics[width=5in]{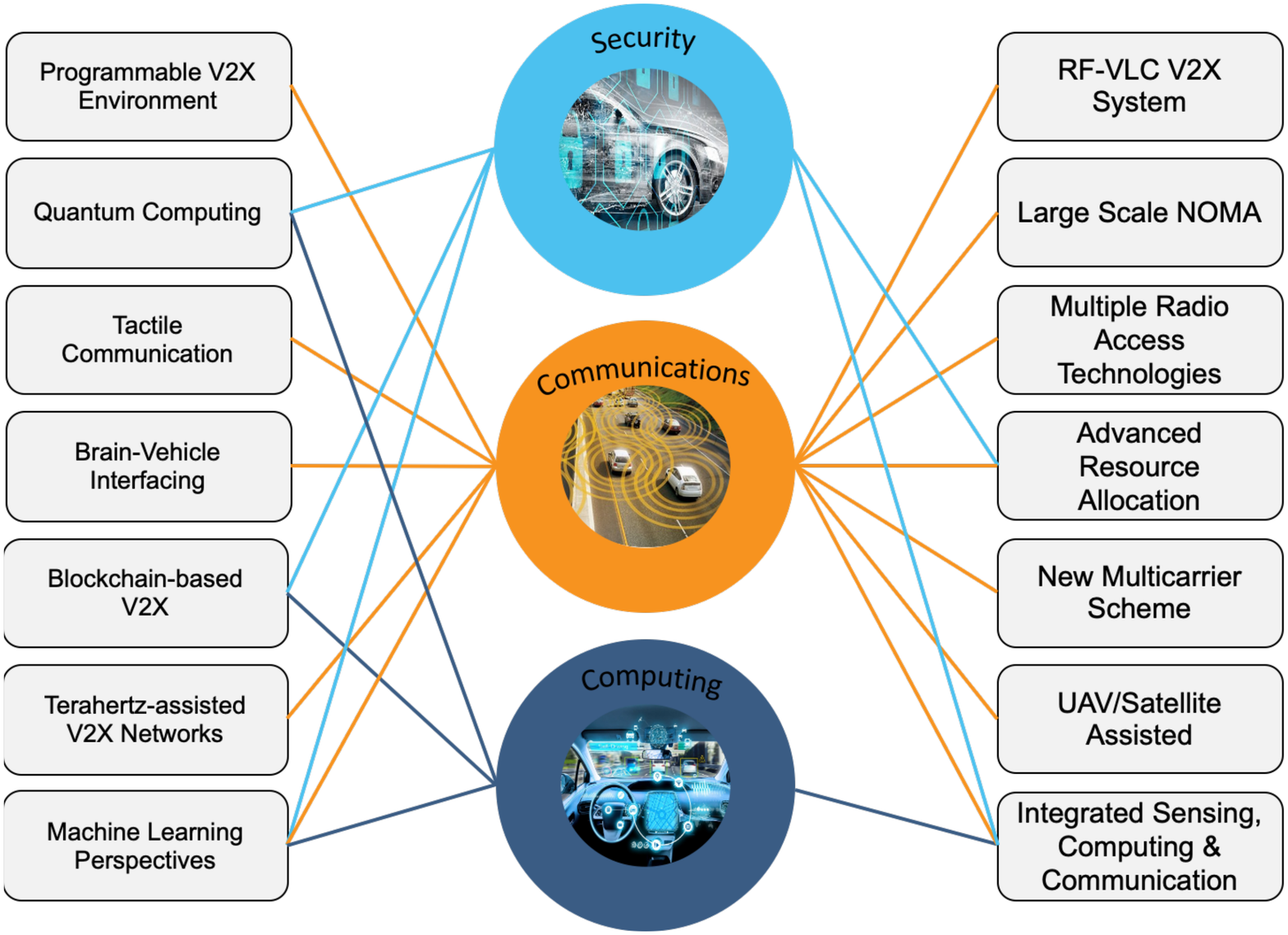}
	\caption{Connection between different technologies and three important aspects of 6G-V2X, namely, communication, computing, and security.}
	\label{fig:taxonomy}
\end{figure*}

\begin{table*}
\caption{An Overview of Key 6G-V2X Technologies}
\label{Table-overview-1}    
\scriptsize
\centering          
\begin{tabular}{|c||P{2cm}||m{3.5cm}|m{4.5cm}|c|P{2cm}|}    
\hline 
Category & Technology &  \hspace{1cm} Strength &  \hspace{1.2cm}  Open Challenges & Maturity & Enhancing Areas \\[0.5ex]  
\hline \hline
\multirow{7}{*}{Revolutionary Tech.} & Programmable V2X Environment &
   \begin{minipage}[t]{\linewidth}
\begin{itemize}[leftmargin=*]
\item Intelligent V2X environments;
\item Effective mitigation of path loss, shadowing, multipath and Doppler effect
\end{itemize}
\end{minipage} &
   \begin{minipage}[t]{\linewidth}
\begin{itemize}[leftmargin=*]
    \item Reflection optimization
    \item Channel estimation in highly dynamic V2X environment
\end{itemize}
\end{minipage}
& $\bigstar\obigstar\obigstar$ & PHY layer,  safety \\ \cline{2-6}
& Tactile Communication &
   \begin{minipage}[t]{\linewidth}
\begin{itemize}[leftmargin=*]
    \item Real-time transmission of haptic information
    \item Enhanced vehicle platooning and remote driving
\end{itemize}
\end{minipage}
&
   \begin{minipage}[t]{\linewidth}
\begin{itemize}[leftmargin=*]
    \item Simultaneous requirements for high rates, ultra-low latency, and high reliability  in high mobility environments
    \item Suitable codecs and efficient reconstruction for the haptic data
\end{itemize}
\end{minipage}
& $\bigstar\obigstar\obigstar$ & Haptic interactions, remote
driving \\ \cline{2-6}
& Quantum Computing &
   \begin{minipage}[t]{\linewidth}
\begin{itemize}[leftmargin=*]
    \item Superior computational capability
    \item Enhanced  security
\end{itemize}
\end{minipage}
&
   \begin{minipage}[t]{\linewidth}
\begin{itemize}[leftmargin=*]
    \item Design of quantum devices
    \item Security architecture, characterization of entanglement distribution
\end{itemize}
\end{minipage}
& $\bigstar\obigstar\obigstar$ & Computing \& security  \\ \cline{2-6}
& Brain-Vehicle Interfacing &
   \begin{minipage}[t]{\linewidth}
\begin{itemize}[leftmargin=*]
    \item Enables brain controlled connected vehicles
    \item Managing uncertainties
\end{itemize}
\end{minipage}
&
   \begin{minipage}[t]{\linewidth}
\begin{itemize}[leftmargin=*]
    \item Scalability of brain-controlled vehicles
    \item Comprehensive real-world testing
\end{itemize}
\end{minipage}
& $\bigstar\obigstar\obigstar$ & Driving experience, disabled
people driving   \\ \cline{2-6}
& Blockchain &
   \begin{minipage}[t]{\linewidth}
\begin{itemize}[leftmargin=*]
    \item Highly distributed
    \item Significantly enhanced security
\end{itemize}
\end{minipage}
&
   \begin{minipage}[t]{\linewidth}
\begin{itemize}[leftmargin=*]
    \item Algorithm design for  ultra-low latency application
    \item Increasing throughput and scalability
\end{itemize}
\end{minipage}
& $\bigstar\bigstar\obigstar$ & Security, safety, EV charging
efficiency  \\ \cline{2-6}
& THz Communications &
   \begin{minipage}[t]{\linewidth}
\begin{itemize}[leftmargin=*]
    \item Extremely high throughput
    \item Higher spectrum
\end{itemize}
\end{minipage}
&
   \begin{minipage}[t]{\linewidth}
\begin{itemize}[leftmargin=*]
    \item Design of transceiver architectures
    \item Propagation measurement and channel modeling
\end{itemize}
\end{minipage}
& $\bigstar\obigstar\obigstar$ & PHY layer, intra-vehicle communication   \\ \cline{2-6}
& ML-aided V2X Design &
   \begin{minipage}[t]{\linewidth}
\begin{itemize}[leftmargin=*]
    \item Suitable for highly adaptive and complex  V2X  environments
    \item Performance enhancement
\end{itemize}
\end{minipage}
&
   \begin{minipage}[t]{\linewidth}
\begin{itemize}[leftmargin=*]
    \item Performing  effective training in highly dynamic environments
    \item Processing big-data in  real-time
\end{itemize}
\end{minipage}
& $\bigstar\bigstar\obigstar$ & PHY \& MAC layers, Security, self driving
vehicle  \\
\hline \hline
\multirow{8}{*}{Evolving Tech.} & Hybrid RF-VLC V2X &
   \begin{minipage}[t]{\linewidth}
\begin{itemize}[leftmargin=*]
    \item Ultra-high data rates
    \item Low setup cost
\end{itemize}
\end{minipage}
&
   \begin{minipage}[t]{\linewidth}
\begin{itemize}[leftmargin=*]
    \item Inter-compatibility between VLC and RF
    \item Interference management
\end{itemize}
\end{minipage}
& $\bigstar\bigstar\bigstar$ &  PHY \& MAC layers\\ \cline{2-6}
& Multiple Radio Access Technologies &
   \begin{minipage}[t]{\linewidth}
\begin{itemize}[leftmargin=*]
    \item Inherent benefits of sub-6 GHz, mmWave and/or THz for long communication range
    \item Hyper-high data throughput
\end{itemize}
\end{minipage}
&
   \begin{minipage}[t]{\linewidth}
\begin{itemize}[leftmargin=*]
    \item Dynamic configurations meet different QoS requirements
    \item Beam and interference management
\end{itemize}
\end{minipage}
& $\bigstar\bigstar\obigstar$ &  PHY \& MAC layers  \\ \cline{2-6}
& Non-orthogonal Multiple Access (NOMA) &
   \begin{minipage}[t]{\linewidth}
\begin{itemize}[leftmargin=*]
    \item Massive connectivity
    \item Ultra-low latency
\end{itemize}
\end{minipage}
&
   \begin{minipage}[t]{\linewidth}
\begin{itemize}[leftmargin=*]
    \item Cross-layer optimization for grant-free NOMA
    \item Adaptive NOMA and OMA
\end{itemize}
\end{minipage}
& $\bigstar\bigstar\obigstar$ &  PHY \& MAC layers  \\ \cline{2-6}
& New Multicarrier Scheme &
   \begin{minipage}[t]{\linewidth}
\begin{itemize}[leftmargin=*]
    \item Significantly enhanced resilience to Doppler
    \item Higher spectrum- and power- efficiencies
\end{itemize}
\end{minipage}
&
   \begin{minipage}[t]{\linewidth}
\begin{itemize}[leftmargin=*]
    \item Backwards-compatibility with LTE and 5G NR
    \item Scalable multicarrier schemes in highly dynamic vehicular environments
\end{itemize}
\end{minipage}
& $\bigstar\bigstar\obigstar$ &  PHY  layer \\ \cline{2-6}
& Advanced Resource Allocation &
   \begin{minipage}[t]{\linewidth}
\begin{itemize}[leftmargin=*]
    \item Cross-layer resource allocation
    \item Context and situation awareness
\end{itemize}
\end{minipage}
&
   \begin{minipage}[t]{\linewidth}
\begin{itemize}[leftmargin=*]
    \item Efficient and scalable deployments
    \item Distributed intelligent solutions
\end{itemize}
\end{minipage}
& $\bigstar\bigstar\obigstar$ &  PHY \& MAC layers  \\ \cline{2-6}
& Integrated Sensing, Localization and Communications &
   \begin{minipage}[t]{\linewidth}
\begin{itemize}[leftmargin=*]
    \item Higher spectral efficiency
    \item Lower hardware cost
    \item Improved situational awareness
\end{itemize}
\end{minipage}
&
   \begin{minipage}[t]{\linewidth}
\begin{itemize}[leftmargin=*]
    \item Unified design of transceivers
    \item New signal processing algorithms
    \item Optimal waveform design
\end{itemize}
\end{minipage}
& $\bigstar\bigstar\obigstar$ & PHY layer, positioning, parking
efficiency   \\ \cline{2-6}
& Satellite/UAV Aided V2X  &
   \begin{minipage}[t]{\linewidth}
\begin{itemize}[leftmargin=*]
    \item Extra wide coverage
    \item Flexible aerial base-station
\end{itemize}
\end{minipage}
&
   \begin{minipage}[t]{\linewidth}
\begin{itemize}[leftmargin=*]
    \item Energy-efficient computation \& transmission
    \item Robust reception in high mobility environments
\end{itemize}
\end{minipage}
& $\bigstar\bigstar\obigstar$ &  PHY \& MAC layers, accident/hazard monitoring \\ \cline{2-6}
& Integrated Computing  &
   \begin{minipage}[t]{\linewidth}
\begin{itemize}[leftmargin=*]
    \item Faster computing and enhanced security
    \item Low operational cost
\end{itemize}
\end{minipage}
&
   \begin{minipage}[t]{\linewidth}
\begin{itemize}[leftmargin=*]
    \item Integration  of cloud, edge, and fog computing
    \item Heterogeneous design  to support different data sources
\end{itemize}
\end{minipage}
& $\bigstar\bigstar\bigstar$ & Computing, efficient navigation \\ \cline{2-6}
  & Integrated  Control and Communications &
     \begin{minipage}[t]{\linewidth}
  \begin{itemize}[leftmargin=*]
    \item Control-communications co-design
    \item Enhanced platooning
\end{itemize}
\end{minipage}
&
   \begin{minipage}[t]{\linewidth}
  \begin{itemize}[leftmargin=*]
    \item Derivation of fundamental limits
    \item Understanding of  control and wireless networks interaction
\end{itemize}
\end{minipage}
& $\bigstar\bigstar\obigstar$ & Control, vehicle platooning  \\ \hline
\end{tabular}
\end{table*}

\subsection{Key 6G-V2X Technologies}
To achieve the aforementioned ambitious goals,
6G will require the integration of a range of disruptive technologies including more robust and efficient air interfaces, resource allocation, decision making,
and computing. Fig. 2 illustrates such a 6G-V2X system where a range of vehicular communication technologies are adopted
to serve various advanced use cases. For example, UAVs and low earth orbit satellites can support V2X systems with
significantly enlarged and seamless coverage, helping enhance the communication QoS particularly in certain blind spots which might exist in traditional terrestrial communication systems. Edge/fog computing and caching will help V2X communication
devices achieve faster computation, optimized decisions, and longer battery life. Visible light communication (VLC) aided V2X
communications will operate along with traditional RF-based communications to achieve ultra-high data rates, low setup cost, low power consumption, and enhanced security.

There are a range of key technologies that we believe will enable the future vision of 6G-V2X as an intelligent, autonomous, user driven connectivity and service platform for ITS. We will introduce these technologies in the following with more detail in the subsequent sections.
We classify these technologies into two categories: \textit{revolutionary} V2X technologies and \textit{evolutionary} V2X technologies. Strength,
open challenges, maturity, and enhancing areas of these technologies are summarized in Table~I. First, we consider technology areas such as intelligent reflective surfaces (IRSs), a range of new ML techniques, and brain-vehicle interfacing as key enablers
(from the perspectives of new materials, algorithms, and neuroscience, respectively) for more intelligent V2X, which
will further enhance and revolutionize evolving V2X technologies. Tactile communication will provide drivers and passengers with an unprecedented travel experience in the future by exchanging sensory information such as haptics anytime and anywhere. Emerging quantum computing technology will endow 6G-V2X systems with superior computational
capabilities, while we will also see significantly enhanced security along with the use of blockchain technologies. Furthermore, terahertz (THz) communications will enable ultra high data rates never experienced before.


{Here, the various technologies listed  in Table I may interact with each other to deliver unprecedented driving experience, especially for fully autonomous vehicles. First, many gigabit per second data rates can be enabled via millimeter wave communication, VLC, and THz communications, whereas ultra-low latency and reliable information exchange can be supported by multiple radio access technologies, new multi-carrier scheme, and advanced resource allocation. For massive and ubiquitous vehicular access, NOMA and satellite/UAV aided V2X are two promising wireless paradigms. Additionally, integrated sensing, localization and communication will contribute to cm-level positioning and cm/s-level velocity estimation accuracy; artificial intelligence and brain-vehicle interfacing will lead to augmented awareness of both the complex physical and electromagnetic environments.

For example, NOMA has been employed in \cite{Tactile2019} to support low-latency tactile internet for autonomous vehicles, thanks to its enabling of massive connectivity and thus faster random access. In NOMA-based V2X, data security is a real issue as one user's decoding often involves the decoding of other users. Such an issue may be addressed by utilizing blockchain for secured networking, as reported in \cite{BlockChain-NOMA}. On the other hand, the low latency nature of NOMA may help tackle the slow access problem in blockchain based V2X networking, imposed by the inherent algorithm in attaining consensus among decentralized entities such as vehicles and infrastructure. Also, to realize both the orthogonal multiple access (OMA) and sparse-code multiple access (SCMA) \cite{SCMA2014} (one of the typical code-domain NOMA schemes \cite{Liu2020}) according to the requirements of specific V2X use cases,  a highly flexible and scalable multi-carrier system is preferred. The selection of new multicarrier waveform for OMA or SCMA and with respect to different QoS requirements (e.g., mobility, spectral efficiency, error rate performance) is of strong interest for reliable, ubiquitous, fast, and efficient wireless access in 6G V2X. Enhanced integrated sensing, localization and communication for autonomous vehicles may also be carried out with the aid of IRS and THz mapping \cite{6GWhite2020,Henk-conf-2021,Henk-mag-2021}.}

{Additionally, enhanced vehicular message dissemination is also expected to transform the future of electric vehicles (EVs), which are the automotive industry's response to address fossil fuel depletion and environmental pollution \cite{ Elghanam2021}. It is anticipated  that 6G V2X will significantly improve both the driving efficiency and battery efficiency of EVs. As the road conditions and the best routes are predictable, optimized driving modes for EVs can be applied for battery and travel time-saving \cite{Gilbert2021,Chau2017}. With 6G V2X, the EV battery status can be better monitored and configured through cloud-based computation or machine learning \cite{Wang2021,Weihan2021}. EV charging times can also be optimized as 6G V2X can guide an EV to find the most convenient charging stations at different periods of a day/season. In this case, a nearest charging station may not be the best as there might be many EVs queuing for charging and also the driving time to that charging station may be long during peak-hour periods. 
}

\section{Revolutionary Technologies  for 6G-V2X} \label{sec:revtech}
In this section, we introduce some of the promising revolutionary technologies with the potential to be used in 6G-V2X.

\subsection{Programmable V2X Environment}
\begin{figure}[htbp]
	\centering
		\includegraphics[width=\figwidth]{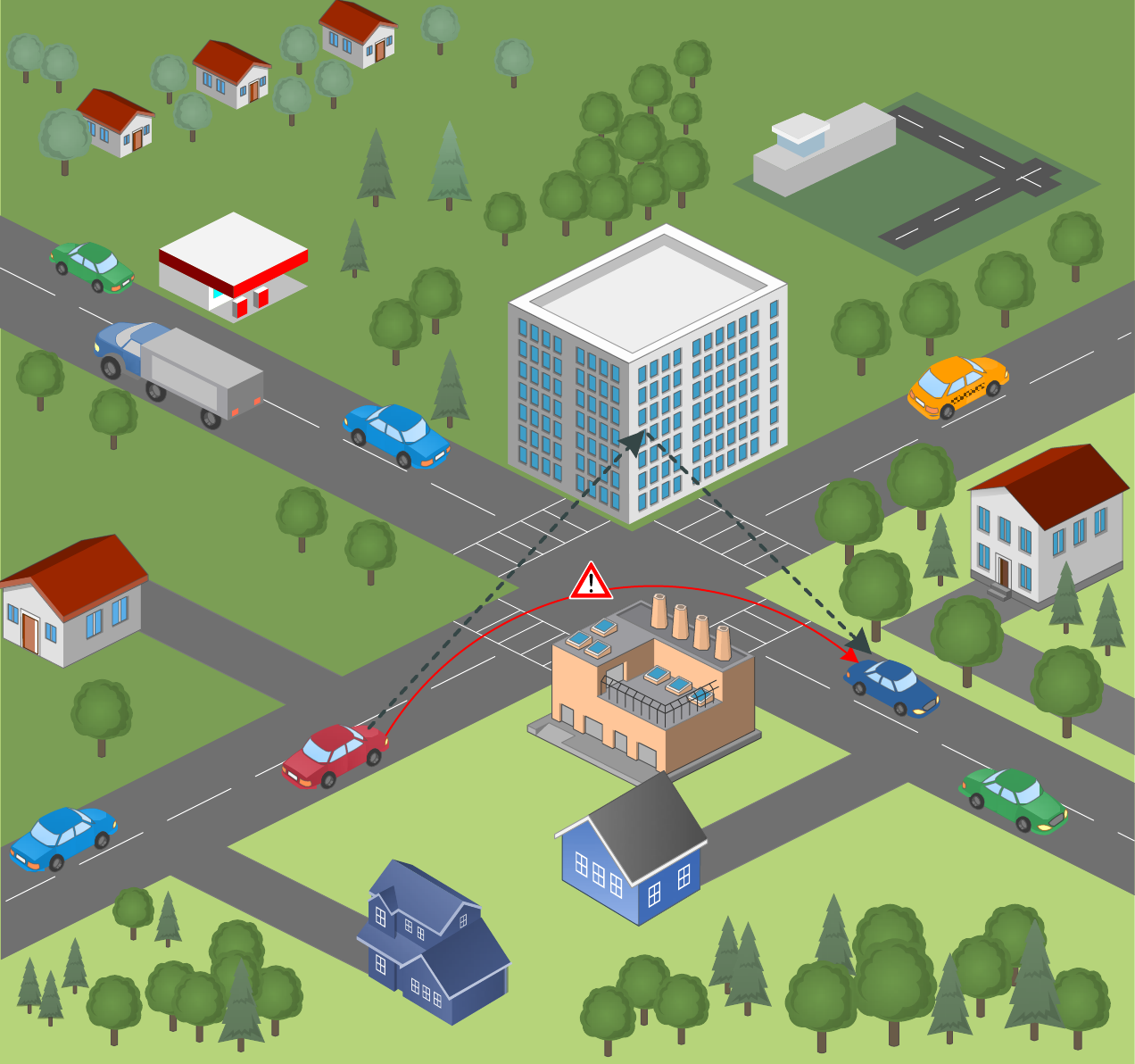}
	\caption{Intelligent reflecting surface at intersection.}
	\label{fig:RIS}
\end{figure}
In conventional communication theory, 
wireless channels are regarded as a destructive and adversarial entity, yielding distorted received signals, causing excessive training overhead, yet having limited channel capacity. One of the most destructive wireless channels is called doubly selective channel (i.e., time-and-frequency selective) which widely appears in vehicular communication systems due to the mobility of transmitter and/or receiver. Furthermore, V2X communication signals may experience significant shadowing effects due to high-rise buildings in urban areas or hills and major vegetation in rural environments. To achieve robust information flow in high mobility channels, LTE and 5G NR-based V2X systems are configured with large  subcarrier spacing\footnote{A relatively large subcarrier spacing is needed in high mobility channels due to the frequency dispersion incurred by the Doppler spread; or the V2X communication system could suffer from substantial increase of inter-carrier interference which in turn leads to drastic deterioration of the error rate performance. Such a phenomenon has been highlighted in \cite{Naik19} where NR V2X with subcarrier spacing of 60 kHz enjoys considerable gains at high velocities (280, 500 kmph) compared to that with subcarrier spacing of 15 kHz.} as well as dense pilot placement. However, this may result in a considerably reduced spectral efficiency,  not to mention very complex signal processing algorithms at the receiver.

More recently a disruptive communication technique called intelligent reflective surfaces (IRSs) \cite{Wu2020,Basar2019}  
has attracted increased attention, aiming at creating a smart radio environment by customizing the propagation of radio wavefronts. Specifically, IRSs are man-made programmable metasurfaces (consisting of a vast amount of tiny and passive antenna-elements with reconfigurable processing networks), which can effectively control the phase, amplitude, frequency, and even polarization of the incident wireless signals to overcome the negative effects of natural wireless propagation. As such, IRSs allow the environment itself to be considered as an element of the communication system, whose operation can be optimized to enable higher rates, enlarged coverage and uninterrupted connectivity.  The recent work in \cite{Ntontin2019} shows that sufficiently large IRSs can outperform traditional relay-aided systems in terms of throughput, while allowing nearly passive implementation with reduced complexity.


6G-V2X can take advantage of IRS in coverage-limited scenarios (e.g., V2X communications operating at millimeter-wave (mmWave) or  THz bands) or unfavorable propagation conditions (e.g., non-line-of-sight communication links).  In such scenarios, the use of an IRS can  enhance the vehicular channel conditions by introducing enhanced multi-path propagation which will result in larger transmission coverage. An out-of-coverage traffic intersection is an ideal use case scenario for using an IRS, because the V2V communication links may be blocked by  buildings and other obstructions. Measurements have shown that the strength of the received V2V signal power reduces quickly over distance away from the intersection due to such blockages \cite{Noor_2018, Noor_2019}.  As such, vehicles  located in perpendicular streets may not communicate with each other very well, which could result in significant degradation of V2V communication performance. To mitigate this issue, IRS may be installed on the surfaces of buildings around the intersection. The communication coverage of transmitting vehicles in the perpendicular streets can thus be enhanced by fine tuning the reflecting elements of IRS. An IRS-assisted vehicular communication scenario is illustrated in Fig.~\ref{fig:RIS}. It is interesting to point out that IRS can be employed for mitigation/suppression of the Doppler effect and multi-path fading, making IRS an appealing research direction for significantly enhanced V2X communication in 6G. {Specifically, recent work \cite{Basar2021} has shown that 1) the rapid fluctuations in the received signal strength due to the Doppler effect can be effectively reduced by using real-time tunable IRS and 2) for more general propagation environments with several interacting objects, even a few real-time tunable IRS can remarkably reduce the Doppler spread and the deep fades in the received signal. For a high mobility wireless channel, a Doppler mitigation method by novel transmission protocol and real-time phase control of IRS has been developed in \cite{Wu2022}.} To efficiently integrate with 6G-V2X, IRS needs to overcome some fundamental challenges such as  reflection optimization, optimal  placement of IRS, channel estimation in a highly dynamic vehicular environment and adaptation to different spectrum ranges.

\subsection{Tactile Communication in V2X}
Tactile communication is a revolutionary technology, which enables a paradigm shift from the current digital content-oriented communications to a steer/control-oriented communications by allowing real-time transmission of haptic or sensual information (i.e., touch, motion, vibration, surface texture) \cite{Sharma_2020}. By integrating human sensual information, tactile communication in 6G-V2X is expected to  provide a truly immersive experience for on-board vehicle users \cite{CalvaneseStrinati2019}.
In addition to traditional applications  of multimedia communications (e.g., on-board meetings/demonstrations, infotainment), tactile communication will enhance  vehicular specific applications such as  remote driving, vehicle platooning, and   driver training by  enabling fast and reliable transfer of sensor data along with the haptic information related to driving experience and trajectories. {Several haptic-based warning signals (e.g., waking up drowsy drivers or catching distracted drivers' attention) have been developed and tested for automotive applications to improve driving safety \cite{4641925, Bazilinskyy2018, Gaffary2018}.   On the other hand, tactile-based V2X can be extremely helpful to vulnerable road users by providing them with appropriate haptic signals that will enhance their safety and activity. For instance, authors in \cite{7385556}   used haptic signals to combine cycling with cooperative driving while supporting cyclists moving in a platoon. The authors observed that the proposed system enhanced cycling behavior without negatively impacting concentration levels.}

Despite tactile communication's immense potential,  there are still many challenges. For example, tactile communication requires extremely high-speed and extremely low-latency communication to ensure reliable and real-time exchange of large volumes of haptic information \cite{Tactile2019}. These stringent connectivity  constraints are very difficult to meet in high mobility vehicular environments. This is because they require higher frequencies (e.g., mmWave or even THz) to meet their data demand. However, those higher frequencies are not very reliable particularly in mobile environments.  For example, in \cite{Chaccour2020}, we showed that even in an indoor environment THz networks may not be able to provide highly reliable high-rate communications. This, in turn, motivates research to develop a new breed of services called highly reliable high rate low latency communications (HRLLC) that can provide a combination of traditional 5G services (e.g., enhanced mobile broadband (eMBB) services that  ignore reliability and URLLC services that ignore rate). Apart from the above challenges, tactile communication poses several fundamental challenges including design of  application-specific control and communication protocols,   development of   human-to-machine interfaces for wireless haptic interactions, and design of suitable haptic codecs to capture and represent the haptic data, and exact  reconstruction of received haptic data.

\subsection{Brain-Vehicle Interfacing}
In a brain-controlled vehicle (BCV), the vehicle  is  controlled by the human mind rather than any physical interaction of the human with the vehicle. {For people with  disabilities, BCV may  offer  great potential  for improved independence by providing an alternative interface for them to control vehicles \cite{Audi}. On the other hand, brain-vehicle interfacing may lead to an improvement in manual driving by predicting a driver's actions and detecting discomfort \cite{Nissan}.} Although the current vision is for fully automated vehicles,  the adaptability of humans will play an irreplaceable role  in  managing  the  uncertainties  and  complexity of autonomous driving \cite{Lu_2020}. By keeping humans in the loop, a BCV is  expected to mitigate the limitations of autonomous driving  in challenging and uncertain  environments such as rural and unstructured areas.   Current wireless communication (e.g., 5G) and computation  technologies are not able to realize BCV as services related to brain-machine interactions will require simultaneously ultra-high reliability, ultra-low latency, and ultra-high data rate communication and ultra-high-speed computation.  For example,  a coarse estimation of the whole brain recording  demand  is  about  100  Gbps \cite{moioli2020neurosciences},  the transmission of which  is  not  supported  by   existing wireless technologies. However, through full-phased brain-vehicle interfacing and  machine learning techniques, 6G-V2X must support the learning  and adapting to the behaviour of human drivers.


{Recently, the feasibility of BCV has been demonstrated in academia (e.g., \cite{Bi_2013,Fan_2015}) as well as by the car industry (e.g., \cite{Audi,Nissan}).} Authors in  \cite{Bi_2013} and \cite{Fan_2015} have shown  a brain–computer interface-based  vehicle  destination  selection  system. Although  successfully  tested  under  different  conditions, the currently designed BCV is  not  a  scalable  solution  since  this would  require  a  wireless  connection  to  support  brain-machine interactions  with high coverage, availability, speed, and low latency to provide reliability  and  safety  for  the  end-users. THz communications can be a potential solution to enable high-throughput and low-latency brain-vehicle interfacing.  Fundamentally different performance metrics (e.g., quality of physical experience (QoPE))  need to be introduced and quantified to capture the physiological characteristics and then map into the conventional wireless QoS metrics \cite{moioli2020neurosciences}. Moreover, extensive real-world experiments are required to demonstrate the effectiveness of  BCVs, as  most of the existing works on BCV has been verified through simulation only.


\subsection{Blockchain-based V2X}
The widespread deployment of V2X networks very much relies on significantly enhanced security for large scale vehicular message dissemination and authentication. The consideration for this imposes new constraints for resource allocation in V2X networks. For example, mission critical messages should have ultra-resilient security to deal with potential malicious attacks or jamming, whilst multimedia data services may require only lightweight security due to the large amount of data. These two types of security requirements lead to different frame structures, routing/relaying strategies, and power/spectrum allocation approaches. 6G-V2X can adopt a blockchain system that is viewed as  a disruptive technology for secured de-centralized transactions involving multiple parties. Compared to  traditional security and privacy techniques, the use of blockchain can  provide a wide range of enhanced security and  privacy services without requiring  any third parties \cite{Nguyen2020}.  Through the  inherent distributed ledger technology of  blockchain, 6G-V2X communication can perform distributed security management,  offloading certain tasks with mobile cloud/edge/fog computing, and content caching. A blockchain-based security solution (e.g.,  smart contract or consensus mechanism) in 6G-V2X is expected to  not only allow verification of the authenticity of a message, but also preserve the privacy of the sender \cite{Kang2019,Yazdinejad2019}. Moreover, the  characteristics of blockchains are of interest for management of unlicensed spectrum, which allows different users to share the same spectrum.   6G-V2X may also utilize a blockchain-based spectrum sharing approach, which has the potential to provide secure, smarter, low-cost, and highly efficient decentralized spectrum sharing \cite{Zhang_VTM_2019}.

 While several attempts have been made to realize a blockchain-based communication network \cite{Nguyen2020}, a straightforward adoption of an existing blockchain technology is not suitable for a V2X communication scenario due to its dynamic network characteristics and real-time data processing requirements. Despite blockchain's great potential in enabling enhanced security and network management, the technology itself suffers from high latency and hence new blockchain algorithms with ultra-low latency need to be developed before they can be applied to 6G-V2X. Limited throughput and scalability of current blockchain technology are also major open problems that require a thorough investigation.

\subsection{Terahertz-assisted V2X Networks}


THz  communication, which operates at  terahertz bands (0.1-10 THz),   is  envisioned as a promising approach to alleviate increasingly congested spectrum \cite{Akyildiz2020, Yuan2020} at lower frequencies.  Leveraging  the availability of ultra-wide bandwidth,  THz communication will be able to  provide transmission rates ranging from hundreds of Gbps to several Tbps.  Such an extremely high throughput will enable a plethora of new V2X application scenarios such as  ultra-fast massive data transfer between vehicles and haptic communications. Since THz communication is able to provide fiber-like data rates without the need for wires between multiple devices at a distance of a few meters, it may also be used in on-board use cases such as the BCV scenario, where extremely high throughput and low latency wireless communication is required.

While the THz spectrum brings a number of unique benefits, there are many major challenges to be addressed, such as transceiver architectures, materials, antenna design, propagation measurement, channel modeling, and new waveforms. In particular, it is essential to characterize and understand THz radio propagation in different V2X scenarios such as highway, urban, and in-vehicle. One of the main challenges  in THz-assisted 6G-V2X will be the effective use of traditional cellular and new THz bands. As such, suitable dynamic resource scheduling is required to exploit their unique benefits. For example, while THz communication offers very high data rates, it is only suitable for short-range V2X communications. In this case, resources may be allocated in THz bands to those transmitters with receivers within a short range. Note that  appropriately designed relaying or IRS techniques (as done in~\cite{Chaccour2020_IRS}) can be potential solutions to extend the coverage of TH-based V2X communications.


\subsection{Quantum Computing Aided V2X}
{Quantum computing  is considered as one of the revolutionary technologies for generic 6G wireless communications in a number of seminal works (e.g., \cite{Tariq2020,Akyildiz2020,Gui2020,Yang2019_6G,Zhang_VTM_2019}). However, the development of practical quantum computing and communication systems are in their infancy and practical solutions may be quite some time off. Therefore, as mentioned in \cite{Saad2020}, quantum computing and communications may potentially play a role towards the end of the 6G development or even beyond for 6G+ technologies. Nevertheless, once some form of quantum computing is available for 6G communications, it can be expected to make its way into V2X applications as well.

If available, we can envision that quantum computing will offer enhanced security in V2X communications. Note that security in V2X communications is significantly more important than in traditional communications since, for example, a security breach in an autonomous vehicle can cause fatal accidents. As the wireless spectrum is shared between vehicles and other types of cellular users (e.g., pedestrians), V2X communications may be vulnerable to malicious attacks, and traditional encryption strategies may not be adequate.  Quantum computing has the inherent security feature of quantum entanglement that cannot be cloned or accessed without tampering with it \cite{Tariq2020}, making it an appropriate technology to enhance 6G-V2X communications security. Moreover, quantum domain security is based on the quantum key distribution (QKD) framework that allows to detect  any malicious eavesdropping attempt. For example, the use of quantum federated learning to securely execute learning tasks among  vehicles can be an important use case~\cite{Chehimi_2021}.

In addition to the  enhanced security feature, the advent of quantum computing promises a radically enhanced computational capability offering to significantly enhance and optimise 6G-V2X services through fast execution of extremely complex and currently time-consuming optimisation algorithms. For example, the implementation of  advanced machine learning algorithms which require big data processing and massive training (e.g., finding an optimum geographic  route with multiple objectives)  is a very challenging task. In such scenarios, traditional computing  often sacrifices optimality, while  quantum computing  can efficiently  achieve optimality with reduced complexity  \cite{Akyildiz2020,Botsinis2019}.}




Although quantum computing is seen as a promising technology,  much more research is required to turn it into a widely usable technology in order to exploit its potential. For example, current quantum computer chips can only operate at extremely low temperature (close to zero Kelvin),  which makes them at best only  usable on the vehicular infrastructure side. To use  them in vehicles, significant research is needed on the thermal stability of quantum computer chips. Other fundamental challenges include development of large-scale quantum computing, design of quantum security architectures, and characterization of entanglement distribution.



\section{Evolutionary Technologies  for 6G-V2X} \label{sec:evotech}
In this section, we present a range of technologies that may be classified as evolutionary. While they have reached a certain maturity due to  extensive research, testing, and deployment in the past, significant further development and trials are needed to adapt them to meet the new challenges and requirements in 6G-V2X.

\subsection{Hybrid RF-VLC V2X System}
\begin{figure}[t]
	\centering
		\includegraphics[width=\figwidth]{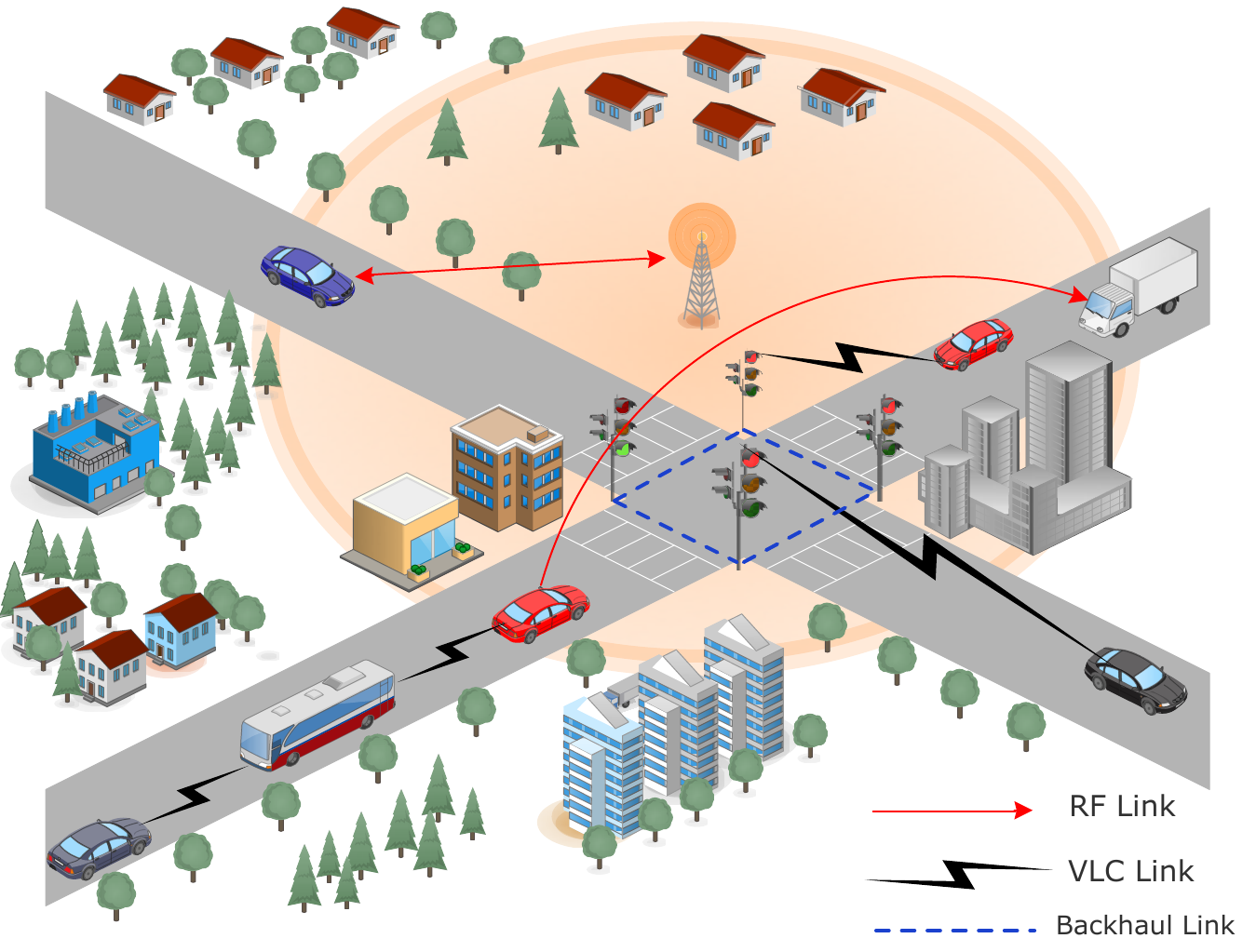}
	\caption{Hybrid RF-VLC-based V2X communications.}
	\label{fig:RF_VLC}
\end{figure}

In 6G-V2X, it is expected that the vehicle and its occupants will be served at extraordinarily high data rates and with extremely low latency. However, this feature may not be feasible with standalone radio-frequency (RF)-based V2X communication as  conventional RF-based vehicular communication often suffers from  high interference,  large latency,  and  low packet delivery rates  in  highly dense scenarios \cite{mecklenbrauker2011vehicular, santa2010analysis}.  One alternative approach may be the combination of RF and visible light communication (VLC)-based V2X communications, where along with radio waves, visible light can be used as a medium of communication in vehicular networks. The ultra-high data rate ( potentially up to 100 Gbps) achieved by   light emitting diode (LED) or laser diode (LD)-based VLC  \cite{ji2014vehicular} and  its inherent features    (such  as  low  power  consumption,  enhanced  security,  and  anti-electromagnetic  interference),  make  VLC  technology an  ideal  candidate  for future   ITS.  Moreover, a  VLC-based  V2X  communication system will require minimum setup cost as VLC-based V2X  can be implemented by using the existing LEDs/LDs in vehicle headlights or pre-installed street/traffic lights.

In V2X networks, VLC can be mainly used in the following three scenarios:  V2V communication through headlights/backlights,  V2X communication through traffic lights, and V2X communication  through street lights. Note that the traffic/street lights can be used to establish backhaul links with one another by using free-space coherent optical communications \cite{Yuan2020}. In addition to enhancing the  data rate, VLC can boost the performance of V2X networks by eliminating the limitations of traditional RF-based V2X communications. For example, in the presence of big vehicle shadowing, RF-based V2V communication suffers from severe packet drop due to high path-loss and packet collision \cite{Hieu_2020}. In this scenario, the transmitting vehicle can communicate with the big vehicle through VLC and then the big vehicle can relay the messages to the vehicles in the shadow region. Similarly, using VLC, traffic/street lights can also be used in the urban intersections to relay the messages  to facilitate communication between vehicles from perpendicular streets, where traditional RF-based V2V communication often suffers from severe packet loss. Note that while RF-based solutions (e.g., big vehicle or roadside unit (RSU) relaying) of the above problems are studied in the literature, such solutions can cause severe interference in the high-density scenarios  due to the RF-based re-transmissions  \cite{Noor_2018}.

Although  extensive research has been carried out on VLC-based V2X communication in the past decade, VLC  has not been included in the 5G-V2X standard. Several open issues still need to be solved for enabling  hybrid RF-VLC V2X. These include interoperability between VLC as  well as RF technologies  and deployment issues. In an outdoor environment, the performance of VLC degrades due to the interference  caused by   natural  and  artificial  light  sources. On the other hand, the received signal strength in VLC may dramatically vary due the vehicles’ mobility \cite{Memedi2020,Pathak2015}.   Hence, ambient lighting induced interference and mobility induced channel variations need to be properly addressed before deploying VLC in 6G-V2X systems.

\subsection{Large Scale Non-orthogonal Multiple Access (NOMA)}
6G-V2X will require massive connectivity for timely, reliable, seamless and ubiquitous exchange of V2X messages. This is to allow connected vehicles to constantly sense and interact with the environments for full situational awareness and hence considerably improved safety. A key enabling technology for 6G-V2X networks to meet these requirements is NOMA. Compared to traditional orthogonal multiple access (OMA) schemes, NOMA allows multiple users to utilize time and frequency resources concurrently for both random access and multiplexing \cite{Cirik2019,Guo2019a}. There are two main types of NOMA: power-domain NOMA \cite{PDNOMA2017,Yuanwei2017} and code-domain NOMA \cite{SNOMA2018,Liu2020}, such as SCMA \cite{LDS2008,Liu2021,Liu2019,SCMA2014} and resource spread multiple access \cite{RSMA2016}, in which multiple users are separated by different power levels and different codebooks/sequences, respectively.
In recent years, NOMA has been proposed for grant-free access to achieve ultra-low latency massive connectivity whilst achieving superior spectrum efficiency.

As a matter of fact, NOMA can be a strong complement to other aforementioned 6G-V2X enabling technologies for use in both V2V and V2I communications. For example, NOMA can be used for distributed V2V autonomous scheduling, where connected vehicles need to contend for the random access control channel to reserve resources for data packet transmissions.  In large scale 6G-V2X networks, the control channel with traditional random access schemes may be saturated, leading to severe collisions on scheduling assignment packets and significant loss of data packet reliability. With the application of NOMA (e.g., SCMA) for control channels, collisions of scheduling  packets can be significantly reduced and the reliability of data packets can thus be improved. NOMA can also be applied in large scale 6G-V2X networks to achieve efficient data packet transmission. This requires multiple communication modes, such as unicast and multicast, which have been added for V2X communications (in addition to broadcast) since Release 15. These multiple communication modes pave the way for the application of NOMA, where broadcast packets may be superimposed by unicast or multicast packets targeting very close neighboring vehicles \cite{Tang2020}.
In this scenario, power domain NOMA, in which a large portion of transmit power is allocated to broadcast packets and the remaining for unicast or multicast packets, may be used.

Despite significant research efforts by both academia and industry, however, NOMA has not been adopted in 5G NR as no consensus has been achieved in 3GPP. While existing research works have been reported on the applications of NOMA to V2X networks, they are mainly focused on the V2I links and centralized resource allocation. Many problems of NOMA remain open. For example, how to efficiently coordinate and schedule different users for NOMA transmissions to co-exist with the current OMA (e.g., orthogonal frequency-division multiple access) communications? How to design a highly flexible and scalable NOMA scheme which can strike a balance between overloading factor, reliability and user fairness? How to design practical and efficient large-scale NOMA for connected and autonomous vehicle (CAV) applications, especially in distributed V2V network scenarios?

\subsection{Exploration of Multiple Radio Access Technologies}

Exploiting the higher frequency spectrum in the mmWave and THz range is vital to achieve the 6G KPIs (e.g., Tbps data rates, billions of connected devices, sub-milliseconds of access latency).
The rich frequency resources at mmWave and THz bands can provide larger bandwidth (e.g., multi-Gigabits and 10s GHz for mmWave and THz, respectively)
than the one available at sub-6 GHz, which is highly congested in existing cellular systems.
These rich frequency resources are needed to enable high data rates and low latency for  6G-V2X communications.
Extensive research has been carried out studying infrastructure-based 5G mmWave communications, such as channel modelling and massive multiple-input multiple-output (MIMO) beamforming.
However, V2X communications in mmWave and THz frequency bands suffer from excessive propagation loss and susceptibility to blockage by obstacles such as vehicles and buildings. In addition, the much smaller cells in mmWave and THz bands may significantly increase the frequencies of handovers.
These problems make it challenging for  mmWave and THz communication systems to provide the relevant QoSs needed for the expected advanced V2X applications \cite{Chaccour2020}.
It is foreseen that multi-radio access technologies with sub-6 GHz, mmWave and/or THz will be needed to work together in future 6G-V2X networks \cite{Coll-Perales2019}. For example, while mmWave and THz radios will provide extra bandwidth and capacity to 6G-V2X networks, but sub-6 GHz radios are critical for the enabling of long communication ranges and connectivity stability.

There are a number of challenges to be addressed for efficient usage of multi-radio access technologies.  From the perspectives of mmWave and THz V2X communications, the excessive propagation loss and signal blockage necessitate the use of directional beamforming. The directional connectivity makes V2V operation with mmWave and/or THz radios very challenging for vehicles in high moving speeds.
Communication between two vehicles over a mmWave link including the physical channel and communications of mmWave for V2V  has been studied \cite{zugno2020standardization}.  However, the challenges that mmWave and THz introduce at the MAC layer due to beamforming communications remain largely open for 6G V2V networks.
Novel schemes for coordination and collaboration among these multi-radio access technologies are needed in order to tackle MAC layer challenges, such as fast link configuration and beam management, contention-based channel access, sidelink autonomous scheduling,
distributed congestion control and interference management at MAC layer. Moreover, the use of IRS combined with high frequencies is worthy of a close investigation as it has the potential to help alleviate some of those challenges, as shown in \cite{Chaccour2020_IRS}.


\subsection{Advanced Resource Allocation}
Radio resource management (RRM) will play a crucial role in 6G-V2X networks, especially for providing the QoS required by advanced V2X applications.
The base stations usually take the main responsibility for RRM in the current cellular V2X networks, which has been widely studied in the literature \cite{Calabrese2018,Olwal2016,She2017}.
However, there are several major challenges for RRM in 6G-V2X networks.
As previously mentioned, 6G-V2X networks will very likely need multi-radio technologies to deliver the expected QoS. The resources of different technologies will need to be taken into account in the RRM decision making. Solutions that smartly use the characteristics of different technologies (e.g., higher rate for mmWave and better reliability for sub-6 GHz) are needed.  Moreover, most current RRM approaches in the literature use either fixed rules, analytical models or supervised learning in relatively low-dimensional scenarios. However, existing 5G RRM research has mainly focused on infrastructure-based communications \cite{8717730,8758996,9046784}.  The expected problem dimensionality will significantly increase for  6G-V2X networks partly due to mmWave deployment and coexistence of multiple V2X use cases with additional direct V2V communications and autonomous resource control operational modes. Moreover, the fast-moving nature of V2X networks and the stringent QoS requirements that must be met to support advanced V2X use cases make the RRM problems more challenging.

In order to address the above challenges of supporting multi-radio technologies and increased algorithm complexity, advanced resource allocation schemes are needed, which could be built with the support of context awareness and cross-layer design.
A hybrid RRM framework can be created, in which both dedicated radio resources and a shared resource pool are allocated to the connected vehicles for V2V and V2I communications. The dedicated radio resources to the individual CAVs ensure a basic but critical level of QoS is guaranteed for various CAV applications. The shared resource pools are then provided to flexibly accommodate the temporary loss of mmWave or THz connections and adapt to fast changing network conditions.
The allocation can be adjusted adaptively according to the QoS feedback and the context. Context awareness of the communication system and the driving environment could be pivotal for cross-layer design of RRM solutions. For instance, in \cite{Semiari2018}, we showed how one could use such smart multi-radio solutions (at mmWave and sub-6 GHz) with context-awareness to provide a reliable video performance at high mobility. To deal with the very large action space and time complexity of RRM and QoS control problems, distributed intelligent solutions over multi-radios that are able to dynamically allocate resource blocks (RBs) and power should be designed following the hybrid RRM framework. Reinforcement learning could be applied to design such intelligent solutions. More discussion on the ML-based resource allocation will be presented in section~\ref{sec:Key_ML}.


\subsection{New Multicarrier Scheme}
6G should provide ultra-reliable high-rate V2X communications in high mobility environments. Connected vehicles and high-speed trains, moving at speeds of 1000 km/h or even higher \cite{Akyildiz2020,Zhang_VTM_2019}, will communicate with each other and the surroundings including various sensors, infrastructure nodes (e.g., roadside units, base stations, robots), satellites, and the internet cloud. {Such high speeds may lead to significantly reduced channel coherence time and as a result, the vehicular channel fading coefficients are rapidly time-varying. Considering the effects of large Doppler and multipath propagation, innovation is needed to cope with the increased interference seen by the receiver} \cite{Fan2015}. In both LTE and 5G NR, OFDM and its variants are adopted for high-rate transmissions \cite{Zhang2016,Lien2017}. Nevertheless, OFDM is very sensitive to the Doppler effect which may destroy the multi-carrier orthogonality and result in increased amounts of  inter-carrier interference and inter-symbol interference.  To overcome  this drawback, some advanced multicarrier waveforms \cite{FBMCbook} may be excellent candidates to 6G V2X. A promising multicarrier waveform is filter-bank multicarrier (FBMC) which enjoys tight spectrum containment as well as relatively strong resilience to carrier frequency offsets and Doppler spreads. These advantages give FBMC a great potential for the support of a diverse range of modern use cases where flexible time-frequency allocations are highly demanded \cite{FBMC2017}.

Recently, orthogonal time-frequency-space (OTFS) has emerged as an effective multi-carrier scheme by spreading each information symbol over a two dimensional (2D) orthogonal basis function spanning across the time-frequency domain \cite{Hadani2017,Raviteja2018,Surabhi2019}. In principle, OTFS is capable of converting a time-varying multipath channel into a relatively static delay-Doppler image of the constituent reflectors. The 2D basis function in OTFS, called discrete symplectic Fourier transform (DSFT), is essentially an orthogonal precoding transform to harvest the diversity gain from time, frequency, and space domains. Thus, it would be interesting to investigate new multicarrier transforms with reference to OTFS for enhanced performance in high mobility environments. Besides, the existing NOMA studies are mostly focused on its application for massive machine-type communications with low mobility and low-rate transmissions. To provide ultra-reliable high-rate massive connectivity (driven by augmented reality/virtual reality (AR/VR) and autonomous driving), it is also promising to study the integration of NOMA (e.g., SCMA) and OTFS (or its variants) to exploit the benefits of these two disruptive techniques.

\subsection{UAV/Satellite Assisted V2X}
Due to the inherent property of  wide area coverage, UAVs can  be used as aerial radio access points in the 6G-V2X network. UAVs can provide different types of services for vehicular users, such as relaying, caching, and computing \cite{Cao2018}.   Particularly, in a highly dense vehicular environment, UAVs can cooperate with the static network infrastructures nodes (i.e., base stations) in managing the wireless network to enhance the user experience. Because of their   nearly unrestricted 3D movement,  UAV can offer a number of unique V2X applications  as a  flying  agent, such as:  a)  providing an advance road accident report prior to  the arrival of rescue team, b) monitoring traffic violations to assist law enforcement agencies, and  c) broadcasting warning about road hazards that  occurred in an area not pre-equipped with an RSU \cite{Menouar2017}. Despite significant advancements in UAV technology, there exist several challenges in the area of UAV-enabled V2X system. For example, it is highly challenging to maintain reliable and high-speed wireless communication between  UAVs and ground vehicles, as mobility of both UAVs and ground vehicles will lead to highly dynamic channel characteristics. While line-of-sight  links are expected for UAV-ground vehicle  channels, systematic measurements and modeling of such channels are still ongoing \cite{7470933}.  Several other key challenges include safety and regulations, seamless integration with existing networks,  and limited battery life of UAVs.

\begin{figure}[t]
	\centering
		\includegraphics[width=\figwidth]{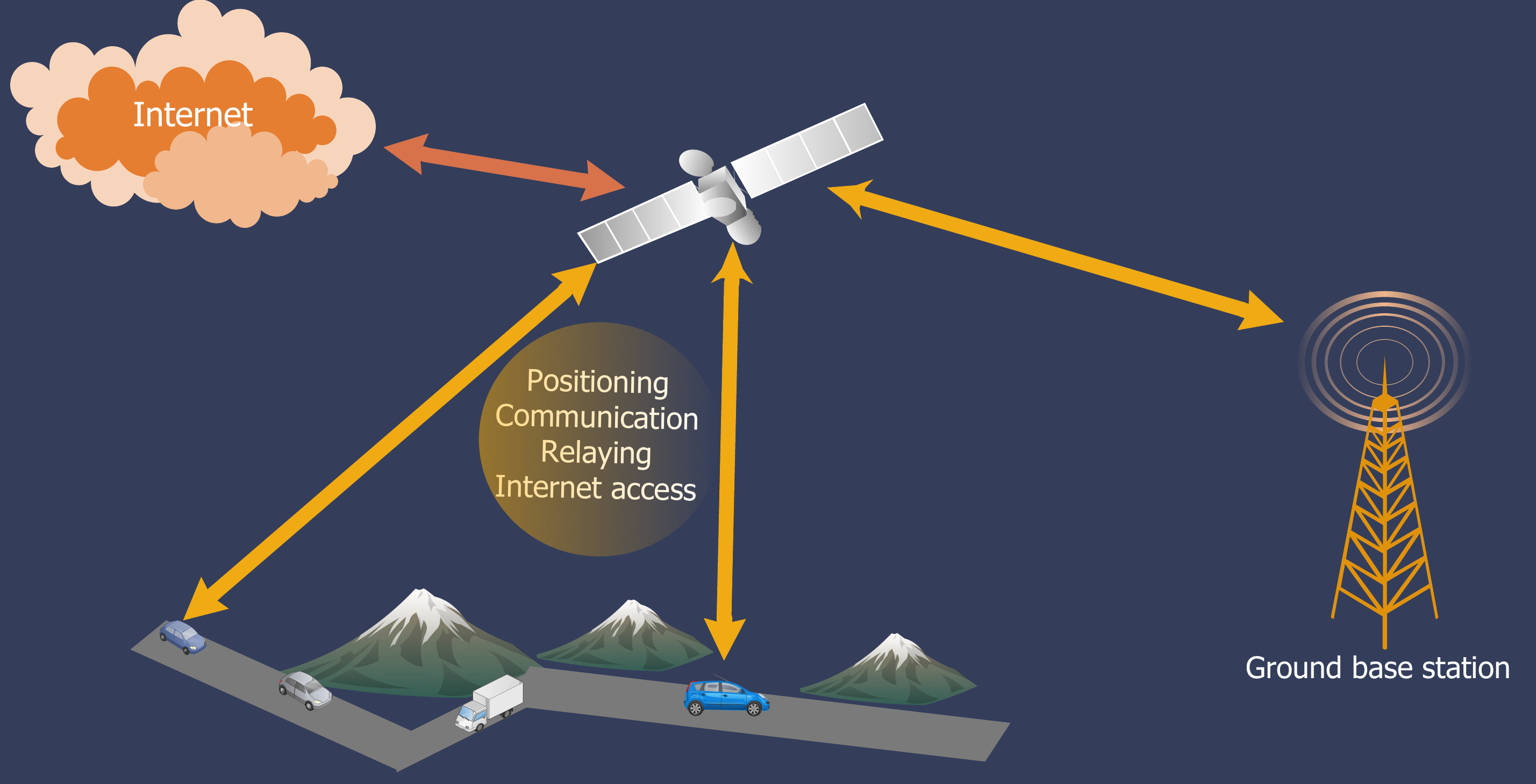}
	\caption{Satellite-assisted V2X communications.}
	\label{fig:Sat_V2X}
\end{figure}

Satellites are another potential aerial communication platform for 6G-V2X communications. An example of a satellite-assisted V2X communication system is illustrated in Fig.~\ref{fig:Sat_V2X}. In current V2X standards, satellites are currently only used for  localization purposes. It is worth mentioning that  the data rates of  satellite communication have been increasing significantly in recent years. For example, multi-beam satellites \cite{Vidal2012} have been widely adopted in satellite communication systems due to their capability to  enhance wireless data rates. Thus, communication via satellite can be a potential technique for 6G-V2X to assist the  communication between vehicle and remote data server in an out-of-terrestrial-coverage scenario. Similar to the UAV-based V2X communication, a satellite can also perform  computing and network management tasks. To enable satellite-assisted V2X communications, an extensive investigation is required to accurately model the characteristics of the channels between satellites and high mobility vehicles. It will also be  challenging to integrate the different communication mechanisms  (e.g., PHY or MAC layer transmission protocol) adopted in V2X and satellite communications.

\begin{figure*}[htbp]
	\centering
		\includegraphics[width=14 cm]{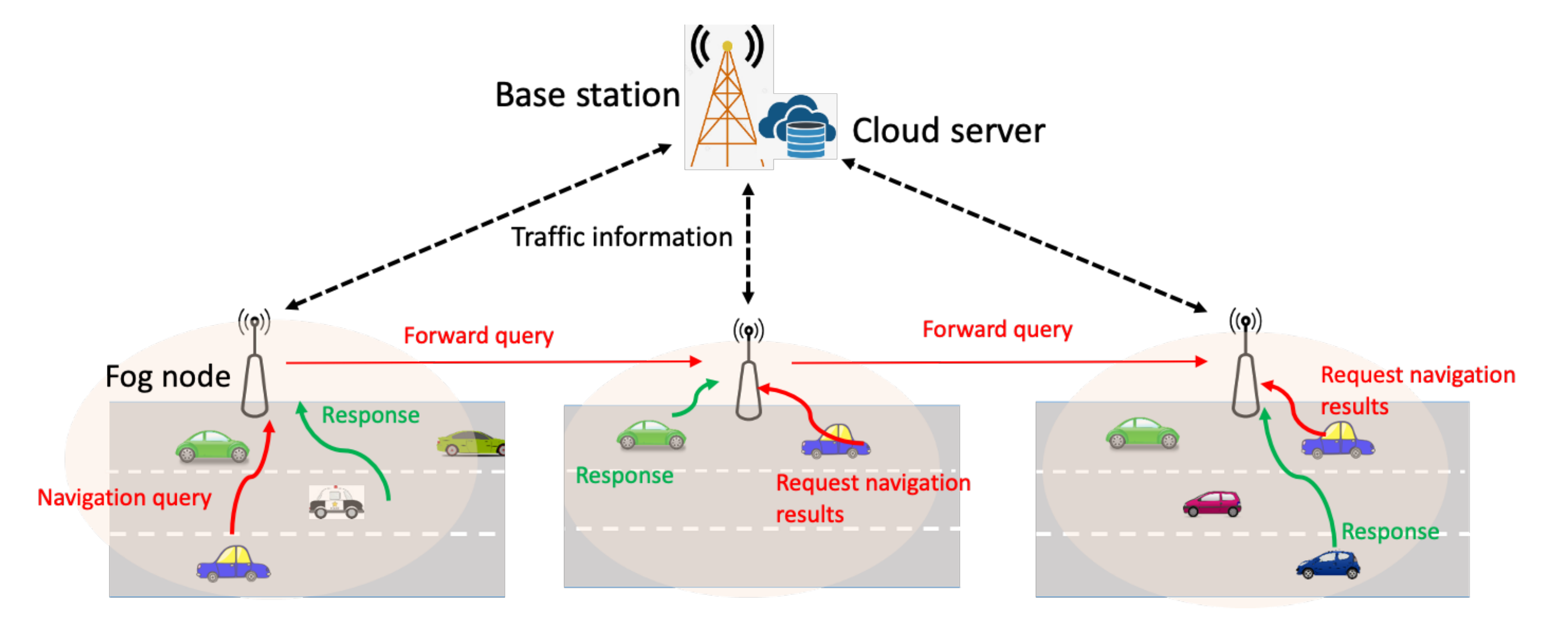}
	\caption{Fog node assisted traffic navigation.}
	\label{fig:Fog}
\end{figure*}

From the PHY point of view, another key research problem is how to attain the highest power transmission efficiency for high-rate and long-range satellite communications. This may not be attained by OFDM due to its high peak-to-average power ratio (PAPR) which ultimately limits its maximum communication coverage. The time has come for the research community to rethink and revisit several traditional modulation schemes which enjoy constant signal envelopes such as continuous phase modulation (CPM) \cite{Anderson1986}. A drawback of conventional CPM is that it may not be suitable to support high-rate communication as OFDM does \cite{Green2011}. Although MIMO could be a way to address this problem, CPM's continuous phase constraint prevents its straightforward integration with MIMO. Recently, a new phase modulation on the hypersphere (PMH) has been developed for load-modulated MIMO \cite{Sedaghat2016,Bhat2018}. It is noted that PMH is capable of achieving the highest power efficiency, while having the advantage of approaching the capacity of Gaussian inputs. Application of PMH for satellite aided long-distance V2X deserves a close investigation.


\subsection{Integrated  Computing for V2X}
Although cloud computing has been widely used in vehicular networks, on their own, cloud-based solutions may not be able to meet many of the very delay-sensitive applications of V2X networks. Edge/fog computing is a newly introduced paradigm, which enables faster distributed  computing and better security at low operational cost. Edge computing operates in a stand-alone mode whereby  the data processing is performed on the nodes that are close to the end users. On the other hand,  fog computing has multiple interconnected layers and could interact with the distant cloud and   edge nodes \cite{Xiaohu2020}.   By leveraging computing resources at  edge/fog nodes  located at the edge of the network, 6G is expected to provide more user aware, scalable and low-latency services for vehicles. Complex algorithms used in V2X network can be solved in real-time  by offloading complex computational tasks to the edge/fog nodes.   One of the use cases of fog computing can be navigation under real-time traffic conditions. For example, fog computing can be used to provide navigation-based real-time traffic conditions as illustrated in Fig.~\ref{fig:Fog}.  The navigation query generated by a vehicle can be sent to the nearest fog node and then relayed to the destination fog node by a hop-by-hop relaying mechanism, where each fog node in the hopping chain collects real-time traffic information in its coverage area.  Upon receiving the traffic reports  from the other fog nodes, the originated fog node computes the optimal path for the vehicle that generated the query.

An integration  of cloud, edge, and fog computing is required in 6G-V2X to exploit the unique benefits of the computing technologies. For instance, together with the edge, fog can perform timely data processing, situation analysis, and decision making at  close proximity to the locations  where the data is generated, while together with the cloud, fog can support more sophisticated applications such as data analysis, pattern recognition, and behavior prediction \cite{Xiaohu2020}. Moreover,  it has been demonstrated that network coding \cite{Ahlswede_2020} can be effectively used to trade abundant computing resources at the network edge for communication bandwidth and latency. 6G-V2X can  exploit the inherent benefits of edge/fog computing and network coding by properly integrating these two techniques. While edge/fog computing provides a number of unique benefits for V2X network, there exist some fundamental  challenges and open problems such as building an integrated computing architecture, handling security and privacy issues, managing handover, and optimising computing resources.



\subsection{Integrated Sensing, Localization and Communication}
{Attaining improved situational awareness for connected vehicles is dependent on not just rapid and reliable communication, but also high-resolution sensing and high-accuracy localization (or positioning/mapping). While sensing helps extract key information (such as ranges, speeds, moving directions, etc) of vehicles, pedestrians, obstacles, and infrastructure, localization permits every vehicle to acquire its precise position which is of utmost importance for safe manoeuvring of vehicles (especially autonomous vehicles). The improved situational awareness will in turn facilitate the exploration of endless opportunities of position-based services as well as advanced V2X applications such as real-time 3D mapping for building an accurate environment model. Like communication, both sensing and localization rely on electromagnetic radio waves to measure, track and interact with the driving surroundings. Hence, it is worthy to explore the natural synergy of sensing, localization and communication by observing the fact that they share similar signal processing operations (e.g., carrier modulation/demodulation, synchronization) as well as hardware implementation. Very recently, in the 6G research community, there has been a strong voice to integrate these three into one converged RF system for higher spectrum- and energy- efficiencies, lower hardware cost/storage, as well as mutually enhanced functionalities \cite{Henk-conf-2021,Henk-mag-2021}.

Among others, there have already been tremendous research attempts concerning the integration of sensing and communication (ISAC) \cite{Liufan2020} and integrated localization and communication (ILAC) \cite{Xiao20}. ISAC, also known as fusion of radar and communication (RadCom) \cite{RadCom2011} and dual-function radar and communication (DFRC), has emerged recently as a disruptive design paradigm for 6G mobile networks with an important application for connected vehicles. For example, in \cite{Kumari2018}, the preambles of the single-carrier 802.11ad system have been exploited to carry out long-range radar sensing in the 60 GHz unlicensed band. It is shown in \cite{Kumari2018} that a gigabits-per-second data rate is achieved simultaneously with cm-level range accuracy and cm/s-level velocity accuracy for a single-target scenario. An architecture for an OFDM based RadCom system has been studied in \cite{RadCom2011} with unique advantages to the radar application, such as, very high dynamic range, independence from the transmitted user data, the enabling of relative velocity estimation, and efficient implementation based on fast Fourier transform. }

Similar to the state-of-the-art of ISAC, centimeter-level localization accuracy is expected by leveraging ultra massive MIMO, mmWave technologies, and UAV/satellite networks. On the other hand, the vehicles' location information can assist wireless networks with a wide range of information such as location-aided channel state information, beam processing, routing, network design, operations, and optimization to effectively utilize network infrastructure and radio resources. Research has already started in this direction. In \cite{Celebi07}, location information has been used in cognitive radios and for network optimization applications, whereas in \cite{Slock12,Dammann13} resource allocation utilizing  location information has been proposed for a multi-user and multi-cell system. A comprehensive survey can be found in \cite{Taranto14} on location-aware communication across various layers of the protocol stack. In \cite{Ghatak18,Jeong15,Destino17,Destino18,Kumar18}, joint localization and data transmission have been studied for 5G networks using different beamforming schemes.  A key challenge here is how to allocate the radio resources effectively between localization and communication while maintaining their QoS requirements. To tackle this problem, ML-based approaches can be used since ML can unfold the full capability of the radio resources intelligently \cite{Xiao20}.  In both ISAC and ILCA, optimal waveform design with ultra-high spectral efficiency is another challenge, which can be solved through effective spectrum sharing techniques or by properly sharing one waveform. The unified design of transceivers also needs to be considered for the seamless integration of localization and communication.

\subsection{Integrated Control and Communication}
Integrated communication and control will play a crucial role in 6G and could potentially help in improving advanced and autonomous V2X services. One of the use cases of integrated communication and control is \emph{vehicle platooning} \cite{7056505,8764451,7466806}, where a group of vehicles travels closely together in a coordinated movement without any mechanical linkage. The key benefits of vehicle platooning include increased road capacity, a rise in fuel efficiency and comfortable road trips. Each vehicle in the platoon must know its relative distance and velocity with its neighboring vehicles in vehicle platooning to coordinate their acceleration and deceleration. Most of the prior works in this direction are  either communication-centric \cite{6766275,9086736,6135795} or control centric \cite{5782989,8025403,8813206}. The former entirely abstracts the control mechanism, while the latter assumes that the performance of the communication networks is deterministic. However, such an assumption can impair the performance of the system. For example, if the exchange of the information is delayed, which can be caused by the uncertainty of the wireless channel, the stable operation of the platoon will be jeopardized.  Therefore, to enable autonomous platooning, integrated communication and control will play an important role. Few studies have been reported in the literature such as \cite{Zeng2019,Gonccalves2020}, which jointly studied the communication and control systems in a V2X network, particularly for vehicle platooning. Although control and communication theories are well studied in the past, existing tools are not yet  adequate for analysing integrated control and communication design \cite{Zhao2019}. For example,  the fundamental limits of wireless control in real-time applications (e.g., vehicle platooning) are still unknown. On the other hand, the tight interaction between vehicle control and wireless networks is not yet well-understood.  Understanding this interaction  will play an important role in the field of integrated communication-control design  for  autonomous vehicles.

\section{Key Machine Learning Perspectives}\label{sec:Key_ML}

Recent advances in ML
research with
the availability of large datasets and storage, and high computational power \cite{Gunduz19Air}
have enabled various novel technologies such as self-driving vehicles and voice assistants\cite{ali20wp6g}.
In view of this background, ML has become
increasingly indispensable and instrumental
towards highly autonomous and intelligent operation of tomorrow's 6G
vehicular networks \cite{Tang20vetNet6G}.

\begin{figure*}[!h]
	\centering
	\includegraphics[trim=0.6cm 2.6cm 0.6cm 7.7cm, clip=true, width=0.75\textwidth]{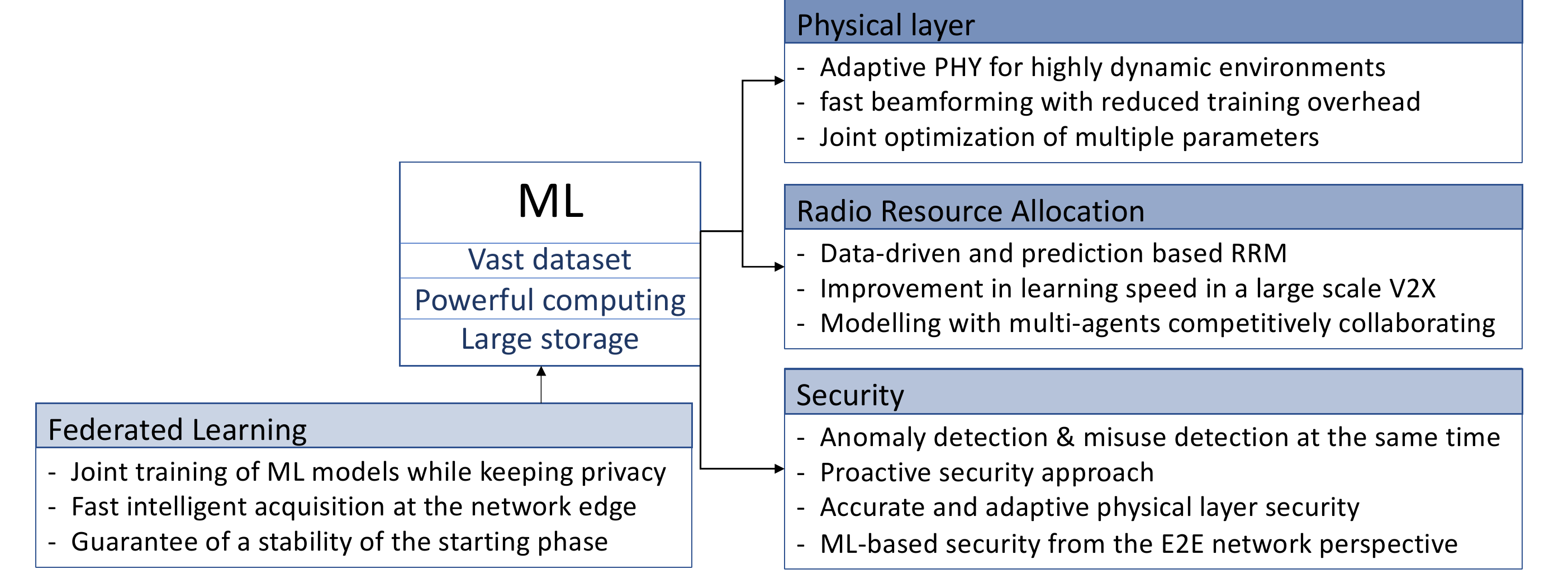}\
	\caption{Summary of potentials and open issues of ML for 6G-V2X}
	\label{fig:summaryRL}
\end{figure*}

The design of traditional wireless communication systems heavily relies on
model-based approaches in which
various building blocks of communication systems
are judiciously modeled based on
analysis of measurement data.
While these model-based approaches have demonstrated their successes in traditional communication system design,
there may be some scenarios in 6G-V2X
in which
accurate modelings (e.g. interference model, accurate channel estimation, etc.) are unlikely  \cite{Gunduz19Air}.
Since ML is capable of extracting the characteristics and identifying (even certain deeply hidden) relationships between input and output data,
it can be adopted as a powerful tool
in the aforementioned scenarios where
traditional communication system design
may suffer from model mismatches.
%
Additionally, the data-driven nature of ML can help  inferences and predictions about channel dynamics, user behaviour, network traffic, application requirements and security threats, thus leading to better resource provisioning and improved network operation \cite{Gunduz19Air}.
{In addition to the key role that ML plays in improving road safety and  driver experiences, the latest progress in  ML techniques is advancing the realisation of the autonomous cars \cite{MLAli2021}. For instance, the data streams of the observed information from cameras, LiDAR's, GPS units, and sensors, can be processed, with which data-driven intelligent decision making takes place through modular perception-planning-action or by end-to-end learning methods for autonomous driving \cite{MLGrigorescu2020}. As far as vision-based ML is concerned, multi-modal reasoning by fusing camera frames and LiDAR scans is investigated in \cite{Prakash2021CVPR} for better object detection to help autonomous driving. While numerous ML-based smart driving applications are expected for future driving, we focus in this paper on the network perspective and highlight the impact of ML in 6G-V2X networks.
}
We discuss the grand vision, significant opportunities, and major challenges of ML, with a key focus on the physical layer, radio resource allocation, and the system security. In addition, we introduce federated learning which is one of the most promising ML technologies. A summary of this section is shown in Fig. \ref{fig:summaryRL}.

\subsection{ML for New Physical Layer}

The vast range of data services in future V2X communications
necessitates the judicious selection of coding, modulation,
waveform, and multiple access schemes.
In 6G-V2X, it is highly desirable to design an adaptive PHY
which can accommodate diverse use cases and the corresponding technological requirements.

The large Doppler spread may cause very rapid channel fading due to the high mobility, preventing accurate channel estimation. Although linear minimum mean square error (LMMSE) estimation produces optimal performance
for linear and stationary channel environments \cite{Savaux17Lmmse},
it may not be effective
for complicated channel conditions especially
in highly dynamic environments. Moreover, since
conventional methods
generally estimate the channel state information first and then recover
the transmit signals, excessive signal processing latency at the receiver may be incurred \cite{Gong19CT}. %
From this point of view, ML is envisaged to help improve channel estimation in future 6G-V2X networks.
In \cite{Li19V2O,Sattiraju20CE,Moon20CE}, deep learning has been adopted to optimize the channel estimation through the
training of neural networks.
However, due to long training period and large training data,
the learning based channel estimation in \cite{Li19V2O,Sattiraju20CE,Moon20CE} includes offline training.
In this case, the potential mismatch between the real
channels and the channels considered in the training phases
could result in performance degradation \cite{ali20wp6g}.
How to carry out effective channel estimation/prediction  in highly dynamic environments is a challenging and interesting research problem.

Furthermore, the design of channel codes such as low-density parity-check (LDPC) or polar codes, for example,
is usually handcrafted with a relatively simple channel model (e.g. Gaussian or Rayleigh).
Such a handcrafted channel code may lead to compromised error correction
capabilities when it is adopted over a high mobility V2X
environment.
In such a scenario, a connected vehicle may experience
a rapid variation of interference when it moves from one
location to another at a high-speed.
Therefore, it is of strong
interest to investigate how to efficiently generate channel code
\textit{on-the-fly} for more flexible rates, lengths, and structures which
are well adaptive to the rapidly-varying communication environments.

As mmWave communication is expected to be widely used in 6G-V2X, beamforming and massive MIMO
technologies are desired to overcome the notorious high path loss problem.
The works in \cite{Va17mW} and \cite{Klautau18mW} pointed out
that the existing beamforming methods for vehicular communication networks suffer from the overload problem  incurred by frequent switching.
Adaptive beamforming in a vehicular communication scenario \cite{Klautau18mW} and \cite{dlMiso}
is possible, but frequent beam training may be needed.
Thus, it is desirable to develop new approaches to help reduce frequent training and heavy overhead while guiding the design of intelligent ML-based beamforming.

The high heterogeneity and dynamics of vehicular networks
will increase the complexity of the environment
including the varying wireless channels.
As discussed, in legacy communication systems,
different blocks of the PHY layer are usually optimized independently for ease of design.
Such a design paradigm may not be optimal when  different QoS requirements
(e.g., latency, reliability, spectrum- and energy-efficiencies, and implementation cost)
have to be met, particularly over very complex vehicular channels.
Different functional blocks at the PHY layer
would need to be configured jointly and adaptively
according to the dynamically varying environment \cite{ali20wp6g}.
For instance, ML-based enhanced adaptive coding and modulation (ACM)  is likely for significantly reduced
communication latency as well as improved robustness \cite{AMCbook,AMC2020}.
ML can also be applied to jointly optimize
multiple configurations.
In this case, an ML-based joint
optimization needs to be developed while taking into account the entire end-to-end physical
layer architecture \cite{Shea17DLp}.

\subsection{ML for Improved Radio Resource Management}

Radio resource allocation, as a classical problem in  wireless networks \cite{Lee14RRM}, has received tremendous research attention in the past years. Although numerous radio resource allocation approaches (e.g., based on greedy algorithm, game theory and optimization) have been investigated, investigations on RRM are needed to satisfy certain new features of 6G-V2X networks such as high mobility, heterogeneous structure and various types of QoS requirements.

First, the mobility of vehicles leads to fast handovers over the links, leading to frequent resource allocation.
While the channel quality and network topology may
vary continuously, conventional resource allocation approaches
would potentially need to be rerun for every small change, incurring huge overhead \cite{Rahim20ra}.
Here again, ML based approaches offer promise as an efficient tool for data-driven decisions to enhance vehicular network performance \cite{Liang19ML}.
For example, in \cite{Tang18CAA}, the prediction capability of ML is adopted to facilitate rapid response to dynamic change of traffic loads.
Their proposed ML-based approach is able to
predict the future traffic load (about bursty traffic patterns)
and assign the available channel to certain links,
thus helping  avoid potential network congestion
as well as rapid channel allocation.

The scarcity of qualified real datasets
for vehicular networks is considered as a big challenge for the use of ML \cite{Morocho19ML}.  %
Reinforcement learning (RL) may be exploited when pre-labeled datasets are not available.
For instance, in \cite{Xu14RL}, RL is adopted for  a vertical handoff strategy of
V2I networks,
in which
RSUs, as learning agents, take into account of information from   vehicles (i.e., average received signal
strength, vehicle velocity and the data type) as well as traffic load for optimal handoff decisions.
Without prior knowledge of handoff behaviour, their  proposed RL-based method can achieve rapid and accurate
handoff to ensure seamless mobility management.
In \cite{Li17RL},
RL is employed for
a user association solution
in heterogeneous vehicular networks (i.e., macro, pico and femto cells).
It is shown that high data rate with load balancing
is attainable by learning
an enhanced association policy based on the data of
traffic loads and the pilot signal strengths received at vehicles.
As indicated in \cite{Xu14RL} and \cite{Li17RL},
RL, which does not require prior knowledge of vehicular environment, is expected to attract increasing research attention  compared to learning approaches
requiring data sets obtained in advance  (i.e., supervised and unsupervised learning).

To use RL for radio resource management problems, it is imperative to seek RL solutions that can quickly converge.
In this regard, one of the major challenges in RL is the so-called exploration and exploitation dilemma \cite{Shen08RL}.
Specifically, an RL agent has to decide between exploration or exploitation, i.e., whether to explore the unknown states and try new actions in search for better ones for future adoption or to exploit those examined actions and adopt them.
While exploration increases the flexibility of the agent to adapt in a dynamic environment at the expense of a possible degradation in the agent's learning accuracy,
exploitation drives the agent’s learning process to local optimal solutions \cite{Yogeswaran12RL}.
When resource allocation problems are modeled with the large state/action space,
finding a good tradeoff between exploration and exploitation is indispensable
in order to improve computational time and convergence speed \cite{SuttonRL}.
For a large-scale network with multiple vehicles, the state and action space in RL may grow very large.
In this case, it is likely that a large number of states are not frequently visited and therefore a much longer time would be required for convergence.

To alleviate this problem, deep reinforcement learning (DRL) has been investigated recently.
In \cite{Ye19dpRL},
multiple parameters of local observations, including channel information of V2V and V2I links and interference levels, are considered to manage the sub-band and power allocation issue. With their problem modeled with a large state/action space, DRL is adopted to extract the mapping relationship between the local observations and the resource allocation-and-scheduling solution.
Moreover, DRL is particularly effective in dealing with the high complexity of  joint optimization problems that often arise when dealing with wireless V2X resource allocation \cite{Ye18AE}.

It is worth mentioning that  a single-agent learning framework is considered in
prior art (e.g., see \cite{Ye19dpRL}),
where
each agent in the same network may take its
action without collaborating with any of the others \cite{Zhao19dpRL}. Such independent choice of actions
could influence other agents' rewards,
hindering the convergence of the learning process \cite{Yau12rl}.
Thus, when ML is applied to a vehicular network with  multiple agents,
the challenge of competitive collaboration should be
considered for effective multi-agent learning.
As an example,
in \cite{Vu20MA},
the problem of joint channel assignment and power allocation in C-V2X networks is studied with multi-agent learning.
When dealing with multi-agent RL, it is often meaningful to use game-theoretic tools \cite{WS19Game} to provide fundamental and rigorous analysis of the RL process.

\subsection{ML for Security Management}

The integration of diverse connectivity and the stringent data provision of services for 6G-V2X will exacerbate the security challenges.
While 6G-V2X aims at providing seamless connectivity between infrastructural nodes and vehicles, the
broadcast nature of vehicular communication makes it susceptible to
malicious attacks.
Various types of malicious attacks (e.g., authentication and authorized attacks, and data forgery and distributions \cite{Sharma20SC}) could target a vehicular network.
Given that, in a V2X system, private user information
such as user identity or trajectory are exchanged over
wireless link, the development of new user
identification and authentication scheme is of particular importance
to maintain secure and legitimate access of data/services/systems \cite{MacHardy18V2X}.

ML can be adopted for detection and prevention of intrusions.
In \cite{Grover11Se} and \cite{Scalabrin17Se},  supervised learning with classification capability is proposed as an effective mean to identify vehicles' abnormal behavior.
It is noted that the training and detection process relies on
existing labeled data, and therefore such a supervised learning may be incapable of detecting
novel/unknown attacks. 
In \cite{Maglaras15Un} and \cite{Sequeira02se},
unsupervised learning which is capable of clustering and does not require labeled data, is considered for real-time detection.
Specifically,
an intrusion detection by using K-means clustering  is proposed for vehicular networks in  \cite{Maglaras15Un}.
To deal with  attacks which can dynamically in real-time,
anomaly detection using unsupervised learning is studied in \cite{Sequeira02se}. However, since these approaches consider either  misuse detection or anomaly detection, they may not be effective
in a real scenario where  known and unknown attacks can  take place at the same time.
In addition,
reactive detection is mainly considered in the existing detection approaches to save communication cost. However,
in a 6G-V2X network where the communication resources are relatively abundant,
proactive exploration-based security approaches are expected to be useful for enhanced security level   \cite{Tang20vetNet6G}.
For example,
in \cite{Khateeb18PD},
a proactive anomaly detection approach is adopted to connected cars for cyber-threat prevention.

Security issues in wireless communication
are usually managed in the upper layers of the protocol stack using cryptographic-based methods.
However, the management and exchange of secret keys will be challenging in  heterogeneous and dynamic V2X networks in which vehicles may randomly
access or leave the network at any time \cite{ElHalawany19PL}.
In this regard,
one can complement standard cryptographic approaches with physical layer security (PLS) solutions \cite{Liu17PLS}.
While PLS techniques exploit the randomness and the physical characteristics of wireless channels
to thwart eavesdropping, these methods can be sensitive to channel
modeling accuracy.
Due to the high mobility and consequently the channel variations  in a V2X scenario, ML can be useful for accurate channel estimation and tracking
which may benefit the design of more effective PLS-based techniques.
Furthermore, depending on the scenarios and services, different security levels are expected.
For example, consider two vehicles which follow each other either on a deserted road or at an busy intersection. Due to the vehicles' movement, the latter has a higher amount of factors which may affect the decision making, resulting in stringent security requirements \cite{furqan20PLS}.
ML may be employed in the latter case to dynamically decide the required security level as well as the most appropriate PLS solution.

ML can also be used to design better control and communication mechanisms that can prevent data injection attacks on vehicular networks, in general, and vehicular platoons, in particular, as shown in \cite{WS19CP}.

When ML is adopted to improve the security, ML-based solution needs to be validated with respect to the end-to-end network performance.
As already mentioned, ML can be used in  functional modules in multiple layers of the networks.
Thus, the use of ML should be synchronized across the network
\cite{Ylianttila20wp} to ensure overall secure communications
\cite{Kumari18E2E}.

\subsection{Federated Learning for 6G-V2X}
\label{Sec:FL}

A critical issue for efficient applications of ML is the training of ML models,
which may be used at the base stations or in the vehicles.
The training of large ML models in remote clouds is an obvious solution but could be time consuming.
One problem is that the fast changing vehicular network and communication conditions
may lead to a slow response to environment changes, thus resulting in degraded performance.
Furthermore, most training samples are generated at the network edges such as base stations and vehicles
and hence the cost and latency of transferring such data to a remote cloud could be very high.
Against this background, local training of  ML models is a desirable solution in 6G-V2X networks.
As each base station or vehicle may hold only a small number of training samples,
joint training of ML models with shared training samples is a potential way to improve ML model accuracy and generalization of performance.
However, a major concern for the joint training is privacy, in which base stations and  vehicles may not want to compromise by sharing training samples.
Federated learning,  emerged in recent years to address
the privacy and communication overhead issues associated to the training of ML models, has attracted extensive research interests for enhanced wireless networks \cite{Park19FL,Chen2020FL,Yang2019,Niknam2020}.

Deemed to be an excellent ML approach for more efficient 6G-V2X networks, there are several technical challenges to be tackled for effective applications of federated learning.
In the existing research works on federated learning of wireless networks, supervised learning is mainly considered.
As reinforcement learning models are more likely to be used,
a scalable federated reinforcement learning framework which can cover many different 6G-V2X use cases is  needed.
In addition, since many V2X applications are mission-critical,
it is often not possible to allow
federated reinforcement learning to learn from scratch to avoid an unstable phase at the beginning of the learning process.
Another challenge of federated learning involving vehicles is the short inter-connectivity between the vehicles.
The vehicles may be out of communication range with the base stations or other vehicles which are involved in the federated learning.
Hence, the vehicles may need to participate in federated learning while they are parked.
Finally, the impact of the wireless channel on the federated learning performance deserves a deeper investigation. As shown in \cite{Chen2020FL}, wireless errors and delays can affect the accuracy of federated learning. This effect can be further exacerbated in a mobile V2X network due to the high-speed mobility of the vehicles and the dynamics of the channel. Further research on the joint design of wireless and learning mechanisms for V2X is needed.


\section{Conclusions} \label{sec:conclu}
In this article,  we have identified a number of key enabling technologies and revolutionary elements of next-generation 6G-V2X networks  by unfolding their potential features and advantages that are far beyond the capabilities of 5G. Further, we have provided an overview of recent advances on applications of machine learning in 6G vehicular networks, which is widely regarded as a key to pave the way towards truly intelligent transportation systems. For each enabling technology, we have highlighted and discussed the major advances, the most pressing challenges as well as the potential opportunities. We expect this article to  provide academic and industry professionals with  key insight into 6G-based next-generation V2X which in turn will stimulate more research with innovative solutions towards the practical design, testing and deployment of these technologies.

\bibliographystyle{IEEEtran}
\bibliography{IEEEabrv,ref_2}

\end{document}